\documentclass[12pt]{article}

\usepackage{epsfig}
\usepackage{amssymb}
\usepackage{graphics}
\usepackage{bbm}
\let\a=\alpha

           \let\p=\pi      \let\r=\rho
  \let\t=\tau      

     \let\L=\Lambda

\def\cA{{\cal A}}

\def\cH{{\cal H}}

\def\cK{{\cal K}}

\def\cM{{\cal M}}
\def\cN{{\cal N}}

\def\cT{{\cal T}}

\def\ie{{\it i.e.}~}

\def\ie{{\it i.e.\ }}

\newcommand{\Zint}{\mathbb{Z}}

\newcommand{\be}{\begin{equation}}
\newcommand{\ee}{\end{equation}}
\newcommand{\bea}{\begin{eqnarray}}
\newcommand{\eea}{\end{eqnarray}}
\newcommand{\ba}{\begin{array}}
\newcommand{\ea}{\end{array}}
\def\nn{\nonumber}

\newcommand{\ft}[2]{{\textstyle\frac{#1}{#2}}}
\newcommand{\Z}{{\mathbb Z}}

\begin{document}
\begin{titlepage}
\rightline{arXiv:0901.0113 [hep-th]} \rightline{ROM2F/2008/26,
ESI-09-2106}
\vskip 2cm
\begin{center}
{\Large\bf (Unoriented) T-folds with few T's}
\end{center}
\vskip 2cm
\begin{center}
{\large\bf Pascal Anastasopoulos}, \ {\large\bf Massimo Bianchi}, \\
{\large\bf Jose F. Morales}\ \large{and} \ {\large\bf Gianfranco Pradisi}\\~\\
{\sl Dipartimento di Fisica and Sezione I.N.F.N.
\\
Universit\`a di Roma ``Tor Vergata''\\
Via della Ricerca Scientifica, 1 \\ 00133 Roma, ITALY}\\
\end{center}
\vskip 3.0cm
\begin{center}
{\large \bf Abstract}
\end{center}

We use the combined action of $\Z_{2}$-chiral reflections (T-dualities)
and  shifts to build ${\cal N}=1,2$ supersymmetric four-dimensional string
compactifications with few moduli. In particular, we consider
 $\Z_2^4$ asymmetric orbifolds of Type IIB on the maximal
torus of $SO(12)$  that mimic  ${\cal N}=2$
Calabi-Yau compactifications with small ``effective'' Hodge
numbers starting from $(h_{11}, h_{21})=(1,1)$. We analyze possible unoriented projections,
providing Type I examples with or
without open strings.  ${\cal N}=1$ oriented 
asymmetric shift-orbifolds of Type IIB with few chiral multiplets are also presented.



\vfill

\end{titlepage}


\tableofcontents

\section{Introduction and summary}

Moduli stabilization in string theory is a long-standing crucial
issue if one is to make contact with low energy (accelerator)
physics \cite{Lust:2008qc}. A very promising path is to turn on
fluxes along the internal directions that generate a potential for
the moduli fields \cite{flux}. However quantization of strings in
the presence of geometric fluxes is only possible for very special
choices involving only open string fluxes
\cite{Antoniadis:2004pp}. Many interesting cases, such as
combinations of NS-NS and R-R closed string fluxes, are only
amenable to an analysis in the low-energy supergravity
approximation. At the same time, a ten-dimensional perspective has
a hard time describing non-geometric fluxes and torsions that may
admit perfectly consistent description in four dimensions as
effective (gauged) supergravities \cite{flux}. Quite remarkably,
resorting to non-geometric constructions, yet based on exactly
solvable (rational) CFT, it is possible to stabilize many if not
all the closed string moduli at tree level or in perturbation
theory. One can then turn on allowed fluxes or invoke perturbative
and non perturbative effects (such as D-brane instantons) to
stabilize the remaining moduli.

Aim of this paper is to further explore controllable mechanisms
(in string perturbation theory) of moduli stabilization based on
exactly solvable (rational) CFT's. We will present simple
non-geometric examples of asymmetric orbifolds
\cite{Narain:1986qm} of special tori
or, equivalently, free fermionic constructions
\cite{Kawai:1986va,abk} with few moduli. The strategy we adopt rests
on the simple observation that chiral (and thus non-geometric)
twists tend to freeze out untwisted moduli while shifts tend to
eliminate twisted ones \cite{Tfolds}. Asymmetric orbifolds of Type
IIB involving chiral twists with no shifts have been previously
studied in \cite{Bianchi:1999uq}.

The simplest non-geometric twist one can think of,  is a $\Z_{2L,R}$
chiral reflection acting on the Left or Right moving closed string
modes. This is nothing but an element of the T-duality group acting on the
worldsheet fields.
 Here we  combine T-duality twists of this type with asymmetric shifts
to build ${\cal N}=2$
compactifications of Type IIB with  few moduli. More precisely, we consider
$\Z_{2L}\sigma_A \times
\Z_{2L}' \sigma_B \times \Z_{2R}\bar{\sigma}_C\times \Z_{2R}'\bar{\sigma}_D$ orbifolds of Type IIB on the maximal $T^6$ torus
of $SO(12)$ with $\sigma$'s some half-shifts.
 We obtain several models with low
``effective'' Hodge numbers starting from  $(h_{11}, h_{21})=(1,1)$.
The construction admits a simple description in terms of free fermions that allows
a systematic search by computer means.

    In view of the possibility
of performing an unoriented projection and including D-branes and
open strings, we mainly focus on non-geometric Type IIB models in
four dimensions with chiral actions on Left-movers mirrored by
identical actions on the Right-movers \cite{BPS, open}.
 ${\cal N}=1$ vacua, following from non-geometric orbifolds  of
 Type IIB  involving   $(-)^{F_R}$ projections breaking all the susy
from the Right-movers will be also considered\footnote{
Being non Left-Right symmetric, these models do
not admit a natural unoriented projection but can be coupled to
generalized D-branes of the kind proposed by one of the authors
in \cite{Bianchi:2008cj}.}.
 In both cases we find ${\cal N}=1$
models with vector multiplets and few chiral multiplets.

Some comments  on the subtle role played by discrete moduli in
asymmetric orbifolds are in order. Asymmetric orbifolds typically
require specific choices of the internal lattice where
``untwisted'' moduli (metric and B-field) are frozen to specific
values. As one is exploring different branches of the original
moduli space, even geometric projections give rise to peculiar
twisted spectra \cite{CRISTINA}. To be specific, starting with the
maximal torus of $SO(12)$ the number of twisted sectors gets
reduced from 48 (16 per each twist) to 12 with a different
chirality structure. As a result, a $\Z_2\times \Z_2$ orbifold of
the $SO(12)$ torus has ``effective'' Hodge numbers $(h_{11}
,h_{21})= (15,15)$ rather than $(h_{11},h_{21}) = (51,3)$  or
$(h_{11},h_{21}) = (3,51)$ as expected when the off-diagonal
components of G and B are set to zero \cite{Berkooz:1996dw}. The
somewhat analogous peculiarities resulting from turning on a
discrete quantized value for the B-field, originally observed in
\cite{BPStor} and then in \cite{MBtor,EWtor,Angelantonj:1999xf},
has been recently reanalyzed in \cite{CBetal, Pesando:2008xt}.

The plan of the paper is as follows. In Section 2 we sketch the
idea of perturbative moduli stabilization by means of (T-duality)
twists and shifts. In Section 3 we describe the basic ingredients
of the free fermionic construction with particular attention to
the case of chiral $\Z_2$ actions. In Section 4 we present the
results of a systematic search over consistent $\Z_2^4$ orbifolds
of Type IIB models with ${\cal N}=2$ susy that admit natural
projections to unoriented ${\cal N}=1$ theories. In particular, we
analyze in some details the ``minimal'' model with ``effective''
Hodge numbers $(h_{11},h_{21})=(1,1)$, that seems to have  escaped
previous scans in the literature \cite{Kiritsis:2008mu,
Donagi:2008xy}. In Section 5 we describe oriented Type II models
with $\cN=1$ susy.
In Section 6,  we present an unoriented model without D-branes based
on the Type IIB model with $(h_{11},h_{21})=(1,1)$ and consistent
with the asymmetric nature of the shift-orbifolds presented in
Section 4. We also analyze a simple instance of an unoriented
model with open strings.
Finally, Section 7 contains our conclusions and some perspectives on the
issue of moduli stabilization. Useful formulas are reported in
Appendices A and B.

\section{Twists and shifts}

In view of moduli stabilization, a particularly promising class of
solvable models are asymmetric orbifolds of tori
\cite{Narain:1986qm}. Indeed, chiral twists tend to freeze out
untwisted moduli while
 (non-geometric) shifts tend to eliminate twisted
moduli. In Left-Right asymmetric constructions level matching
constraints are very demanding and the perspective of a systematic
analysis are daunting. A very simple class of solvable models
which are equivalent to asymmetric orbifolds of special tori are
free fermionic models \cite{Kawai:1986va,abk}. The rules for
constructing modular invariant partition functions compatibly with
both world-sheet and space-time supersymmetry are well understood
and will be reviewed in the next Section. Here we would like to
offer a  geometric interpretation of the free fermion
$\Z_2$-reflections in terms of T-duality twists and shifts.

We will denote by $I_{i}$ a $\Z_{2L}$ chiral reflection of the
$i^{\rm th}$ Left-moving internal bosonic and fermionic
coordinates \bea
 I_i: && X_L^i   \rightarrow  - X^i_L \ ,
 \quad \quad X_R^i
\rightarrow  X^i_R \ , \quad \quad
   \psi^i \rightarrow -
\psi^i \ ,  \quad \quad \tilde\psi^i \rightarrow  \tilde \psi^i \ .
 \eea
 In a similar way one defines the Right-moving twist as
 \bea
 \bar I_i: && X_L^i   \rightarrow   X^i_L \ , \quad \quad X_R^i
\rightarrow -X^i_R \ , \quad \quad  \psi^i \rightarrow
\psi^i \ , \quad \quad \tilde\psi^i \rightarrow - \tilde \psi^i \ .
 \eea
  In addition, we denote by $I_{i_1 i_2\ldots}=I_{i_1} I_{i_2}\ldots $ the simultaneous
  reflections along the
 $(i_1 i_2\ldots )$ directions and similarly for the Right moving ones.
 We will consider $\Z_2^4$ orbifolds with generators including  Left and Right twists
 $I_{3456}, I_{1256}$ and $ \bar I_{3456},\bar I_{1256}$ respectively.
 Each twist breaks half of the Left or Right moving supersymmetries and
one  is left with  $1/4$ of the original spacetime susy.
 Moreover, all untwisted NS-NS moduli fields
 \be |i\rangle_L \otimes |j \rangle_R = \psi^i_{-{1\over 2}} |0\rangle_L \otimes \tilde\psi^j_{-{1\over 2}}   |j \rangle_R
 \quad\quad   i=1,\ldots , 6  \quad,
 \ee
are projected out  by the orbifold group. This implies that both
shape and size deformations of the internal manifold are frozen
out.  Similarly, in the untwisted R-R sector one can see that only
the scalar and the axion that together with the dilaton/axion NS-NS
moduli complete the universal hypermultiplet survive the
projection.

Let us now consider moduli coming from the twisted sector.
 In order to lift as many massless twisted states as possible one
has to combine chiral twists with chiral (non-geometric) shifts.
 We denote the Left moving chiral shift along the $i^{\rm th}$ direction by
 \be
 \sigma_{i}: X^i_L \to X^i_L+\delta \ , \qquad  X^i_R \to X^i_R \quad ;
 \ee
 with $2\delta$ a chiral lattice vector.  Similarly we denote by
 \be
 \bar \sigma_{i}:  X^i_R \to X^i_R+\bar \delta \ , \qquad X^i_L \to X^i_L \quad;
 \ee
 the Right moving shifts and  by  $\sigma_{i_1 i_2\ldots}$, $\bar\sigma_{i_1 i_2\ldots}$ the multiple shifts.
Level matching, \ie modular invariance, puts severe constraints on the allowed  choices of $\sigma$'s.

Another tool one can resort to in order to eliminate twisted
moduli is the judicious choice of discrete torsion
\cite{Vafa:1986wx, Vafa:1994rv}, \ie of the relative signs (for
$\Z_2$)   that multiply orbits of amplitudes not connected
by modular transformations. In the simplest case, discrete torsion
relates the diagonal modular invariant to the charge conjugation
one. More generally, exotic modular invariant combinations of the
chiral characters can change and in some cases drastically reduce
the number of massless combinations.

\section{Free fermions versus asymmetric orbifolds}

In order to perform a systematic search for models with few moduli
in a full-fledged string description we resort to the free
fermionic construction pioneered by Kawai, Lewellen and Tye
\cite{Kawai:1986va} and  by Antoniadis, Bachas and Kounnas
\cite{abk}.

In this description, one fermionizes the internal Left-moving bosonic
coordinates \be \partial X^i = y^i w^i     \quad\quad i=1,\ldots  , 6 \quad ,\ee and rewrites the
worldsheet supercurrent as\footnote{Other choices are possible.}
\be
G = \psi^\mu \partial X_\mu + \psi^i y^i w^i \ , \quad\quad \mu=7,8 \quad.
\ee
All fermions $\{ \psi^\mu,\psi^i, y^i, w^i\}$ are taken to be periodic to start with.
The Right-moving fermions  $\{ \tilde \psi^\mu,\tilde \psi^i, \tilde y^i, \tilde w^i\}$ are introduced in a similar way.

Now, let us consider the orbifolding of the free fermion system by
$\Z_2$ reflections. A reflection is  denoted by a fermion set
$b_\alpha$ that includes all fermions odd under the $\Z_2$.
Spacetime susy and modular invariance put additional constraints
on the allowed fermion sets. Preservation of the worldsheet
supercurrent under parallel transport requires \bea
  \forall i \quad\quad &&   \# ~\psi^i - \# ~y^i -\# ~w^i=0~{\rm mod }~2 \ ;\nn\\
  \forall i \quad\quad &&   \# ~\tilde \psi^i - \# ~\tilde y^i -\# ~\tilde w^i=0~{\rm mod }~2 \ .
  \label{condsusy}
  \eea
 Modular invariance (or level matching) amounts to the following conditions on the basis fermionic sets:
\bea
   n(b_\alpha)&=&0~{\rm mod }~8 \ ;\nn\\
    n(b_\alpha \cap b_\beta)&=&0~{\rm mod }~4 \ ;\nn\\
    n(b_\alpha \cap b_\beta \cap b_\gamma)&=&0~{\rm mod }~2 \ ;\nn\\
       n(b_\alpha \cap b_\beta\cap b_\gamma \cap b_\sigma)&=&0~{\rm mod }~2 \ ;  \label{consistency}
   \eea
    with $n(b)$ denoting the difference between the number of Left- and Right- moving fermions in the set $b$ and the greek indices running
    over the generators of the orbifold group.

The free fermion description of Type IIB on the $T^6$ maximal torus of $SO(12)$ is obtained
by including the following fermionic sets
\bea
F &=& \{ \psi^{1\ldots 8} \, y^{1\ldots 6} \,
w^{1\ldots 6}  | \, \tilde\psi^{1\ldots 8}\, \tilde{y}^{1\ldots 6}\,\tilde{w}^{1\ldots 6}   \} \ , \nn\\
S & =& \{\psi^{1\ldots 8}  \}\ , \quad\quad  \tilde{S} = \{\tilde
\psi^{1\ldots 8}  \} \ . \eea Indeed, the quotient by $F$ results
into a sum over all possible  boundary conditions of worldsheet
fermions, while $S$ and $\tilde{S}$ realize the Left and Right
moving GSO projections, ensuring spacetime susy. Omitting the
integral over moduli space, the resulting partition function can
be written as \bea
{\cal T}_{4_L+4_R} &=& {1\over \eta^2 \bar \eta^2}  |V_8-S_8 |^2 \left( |O_{12}|^2 + |V_{12}|^2 + |S_{12}|^2+|C_{12}|^2\right) \nn\\
  &=& \ft18 \left|{\vartheta_3^4\over \eta^{12}}-{\vartheta_4^4\over \eta^{12}} -{\vartheta_2^4\over \eta^{12}} \right|^2 (  |\vartheta_2|^{12}+ |\vartheta_3|^{12}+
|\vartheta_4 |^{12} ) \ , \label{torus1}
\eea
 where the subscript $4_L+4_R$ reminds that $4_L$ and $4_R$ susy comes from the Left and Right movers, respectively.
We write  partition functions both in terms of characters of
$SO(n)$ at level one or in terms of theta functions, as convenient
in the specific context (see Appendix A for definitions and
conventions). Moreover, signs (discrete torsion) will be chosen
judiciously and respecting the spin-statistic relation.

Another choice consists of keeping only the sets $F$ and $S$, finding the $4_L+0_R$
partition function
\bea
{\cal T}_{4_L+0_R} &=& {1\over \eta^2 \bar \eta^2}  (V_8 - S_8) [O_{12}
\bar V_{20} + V_{12} \bar O_{20} - S_{12} \bar S_{20} - C_{12}
\bar C_{20}]  \nn\\
  &=& {1\over 4 \, \eta^{12}\,\bar \eta^{12}}  \left(\vartheta_3^4-\vartheta_4^4 -\vartheta_2^4  \right)
  (  \vartheta_3^6 \bar \vartheta_3^{10}-\vartheta_4^6 \bar \vartheta_4^{10}
  -\vartheta_2^6 \bar \vartheta_2^{10}  ) \ . \label{torus2}
\eea
 In the following, we will consider asymmetric $\Z_2$ orbifolds of the
${\cal N}=4_L+4_R$  and ${\cal N}=4_L+0_R$ models.  The $\Z_2$ elements will be built out
of chiral reflections $I_i,\bar I_i$ and shifts $\sigma_i,\bar \sigma_i$. In the fermionic language twists
and shifts correspond to the following actions on the worldsheet fermions
\bea
I_i: && \psi^i \to -\psi^i \ , \quad \quad  y^i \to -y^i  \ ;  \nn\\
\sigma_i: &&   y^i \to -y^i \ , \quad \quad  w^i \to -w^i  \ ;
\eea
 with identical expressions for $\bar I_i$ and $\bar \sigma_i$ with fermions replaced by tilde ones.
Alternatively, one can denote reflections and shifts by their associated fermionic sets
\bea
I_i &=& \{\psi^i \,y^i \} \ ,  \quad \quad \sigma_i=\{y^i \,w^i \} \ ,  \nn\\
\bar I_i &=& \{\tilde \psi^i\,  \tilde y^i \} \ , \quad \quad \bar\sigma_i=\{\tilde y^i \,\tilde w^i \} \ .
\eea
 Notice that the non-trivial intersections between the sets are
 \bea
I_i\cap I_j &=& \sigma_i \cap \sigma_j=2 \,\delta_{ij} \ ,  \quad \quad  I_i \cap \sigma_j=\delta_{ij} \ , \nn\\
\bar I_i\cap \bar I_j &=& \bar \sigma_i \cap \bar \sigma_j=2 \,\delta_{ij} \ , \quad \quad  \bar I_i \cap \bar \sigma_j=\delta_{ij} \ .
\eea
  These relations can be used to check the consistency conditions (\ref{consistency}) of an orbifold group generated by
  twists and shifts.

\section{ Models with $\cN=1_L + 1_R$}

\label{sect11}

We performed a systematic search of models with basis sets
$F,S,\tilde{S}$ together with four additional sets of the form
\bea &&
b_1 =(b_{1L},b_{1R})= I_{3456}\, \sigma^{i_1 i_2 \ldots }\,\bar \sigma^{k_1 k_2 \ldots } = \{ (\psi \,y)^{3456}  \, (y\, w)^{i_1 i_2 \ldots }  | (\tilde y\, \tilde w)^{k_1 k_2 \ldots }  \} \ , \nn\\
&&b_2 =  (b_{2L},b_{2R})=I_{1256}\, \sigma^{j_1 j_2 \ldots }\,\bar \sigma^{l_1 l_2 \ldots } = \{(\psi\, y)^{1256}  \, (y\, w)^{j_1 j_2 \ldots }  | (\tilde y\, \tilde w)^{ l_1 l_2 \ldots }   \} \ , \nn\\
&&\bar{b}_1 =(b_{1R},b_{1L})= \bar I_{3456}\, \sigma^{k_1 k_2 \ldots }\,\bar \sigma^{i_1  i_2 \ldots }  = \{ ( y\,  w)^{k_1 k_2 \ldots }  | (\tilde\psi\, \tilde y)^{3456}
  (\tilde y\, \tilde w)^{i_1 i_2 \ldots }     \} \ , \nn\\
&&\bar{b}_2  =(b_{2R},b_{2L})= \bar I_{1256}\, \sigma^{l_1 l_2 \ldots }\,\bar \sigma^{ j_1  j_2 \ldots } = \{ ( y\,  w)^{ l_1 l_2 \ldots } | (\tilde\psi\, \tilde y)^{1256}
(\tilde y\, \tilde w)^{j_1 j_2 \ldots }  \} \ , \eea
 The scanning ran over all choices of sets  $ (i_1  i_2 \ldots) $,  $(j_1 j_2 \ldots) $,   $ (k_1  k_2 \ldots) $,  $(l_1 l_2 \ldots) $ compatibly with the conditions (\ref{consistency}).
  Each set $b_\a$ breaks half of the spacetime susy's arising from
the Left- or Right- moving sector. One is thus left with $\cN =
1_L + 1_R$ susy.

 Defining for notational convenience\footnote{The product is defined as $b_i b_j =  b_i \cup b_j  -  b_i \cap b_j$} $b_3=b_1 b_2$, $\bar b_3=\bar b_1\bar b_2$,
       the generic orbifold group element can be written as $b_{a} \bar b_{ b}$
   with $a,b=0,..,3$  and $b_0=\bar b_0=1$.
     We recall that a contribution of a single Left moving fermion among $\{\psi^i,y^i,w^i\}$ is given by
     $\left( {\vartheta_s / \eta}\right)^{1\over 2}$ (with
     $s=2,3,4$ labelling the spin structure), and similarly for Right moving fermions with $\vartheta_s / \eta$ replaced by  $\bar\vartheta_s / \bar\eta$.
     The $\Z_2$ actions are thus equivalent to
  \bea
\Z_2: 
  &&      \vartheta^{1\over 2}_2\to   \vartheta^{1\over 2}_1    \quad ,\quad  \vartheta^{1\over 2}_3\leftrightarrow \vartheta^{1\over 2}_4 \quad ,\quad
     \bar \vartheta^{1\over 2}_2\to  \bar \vartheta^{1\over 2}_1 \quad ,\quad  \bar \vartheta^{1\over 2}_3\leftrightarrow \bar \vartheta^{1\over 2}_4 \ .
  \eea
The torus partition function can then be written as
\be {\cal T}=\ft{1}{16 \ \eta^8 \bar \eta^8}  \sum_{a,b,c,d=0}^3\, \rho_{ac}\, \bar
\rho_{bd}\,   {\bf \L}[^{a b}_{cd}]  \quad ,    \ee where
${\bf \L}[^{ab}_{cd}]$ denotes the contribution of the
 $(b_c \bar b_{d})$-projection in the $(b_a \bar b_{b})$-twisted sector,
i.e.
\be
{\bf \L}[^{ab}_{cd}]=\ft12 \epsilon_{a,b,c,d} \sum_{\alpha,\beta=0,{1\over 2}} \prod_{i=1}^{12}
 \ \vartheta[^{\alpha+b_{aL,i}+ b_{bR,i}}_{\beta+b_{cL,i}+ b_{dR,i} }]^{1\over 2}
 \ \bar{\vartheta}[^{\alpha+b_{aR,i}+ b_{bL,i}}_{\beta+b_{cR,i}+ b_{dL,i} }]^{1\over 2}
\ee
 with $i$ running over the 12 lattice fermions $y^i w^i$ and $b_{La,i}$,$b_{Ra,i}$   being $0$ or $\ft12$ depending on whether the $i^{th}$ fermion is even or odd under $b_{aL}$ and $b_{aR}$ respectively.   $\epsilon_{a,b,c,d} $ are signs fixed by modular invariance up to discrete torsions.
The precise form of ${\bf \L}[^{ab}_{cd}]$ depends on the
 details of the specific model.
Finally the $\psi$-contribution to the amplitudes can be written as \bea
\r_{00}&=&- \frac{\vartheta_{2,{\rm st}  } \vartheta_2^3-\vartheta_{3,{\rm st} } \vartheta_3^3 +\vartheta_{4,{\rm st} } \vartheta_4^3}{2 \eta ^4} \ , \nn\\
\r_{0h}&=&\frac{ \vartheta_{3,{\rm st} } \vartheta_3 \vartheta_4^2-
\vartheta_{4,{\rm st} }\vartheta_4 \vartheta_3^2}{2 \eta
   ^4} \ , \nn\\
\r_{h0}&=&\frac{ \vartheta_{3,{\rm st} } \vartheta_3 \vartheta_2^2-
\vartheta_{2,{\rm st} }\vartheta_2 \vartheta_3^2}{2 \eta
   ^4} \ , \nn\\
\r_{hh}&=&\frac{ \vartheta_{2,{\rm st} } \vartheta_2 \vartheta_4^2-
\vartheta_{4,{\rm st} }\vartheta_4 \vartheta_2^2}{2 \eta
   ^4} \ , \nn\\
\r_{13} &=&\r_{23}=\r_{32}=-\r_{12}=-\r_{21}=-\r_{31}=\frac{i
\vartheta_{1,{\rm st} }\vartheta_2 \vartheta_3 \vartheta_4} {2
\eta ^4} \ , \eea with $h=1,2,3$ and the subscript  ``st'' denoting the
contribution coming from the spacetime part that encodes the
helicity of the particle (see Appendix A for the definitions of
the amplitudes given in terms of the $SO(2n)$ characters). The
massless content of each model can be read by plugging in the
partition function
 the well-known theta expansions
\bea
 \vartheta_{1,{\rm st}} &=& (S-C) q^{1\over 8}+\ldots \ ,  \quad\quad
 \vartheta_{2,{\rm st}} = (S+C) q^{1\over 8}+\ldots \ ,  \nn\\
  \vartheta_{3,{\rm st}} &=& 1+V\, q^{1\over 2} \ldots  \ , \quad\quad\quad ~~~
   \vartheta_{4,{\rm st}} = 1-V\, q^{1\over 2} \ldots     \ ,  \nn\\
 \vartheta_2 &=& 2 q^{1\over 8}+\ldots  \ , \quad\quad\quad\quad ~~~~
  \vartheta_{3,4} = 1+2\, q^{1\over 2} \ldots  \ , \, \quad\quad\quad ~~   \eta = q^{1\over 24}+\ldots \ .
\eea
where $O,V,S,C$ denote a four-dimensional scalar, vector, left spinor and right spinor, respectively.
 The result can always be written in the form
 \bea
 {\cal T}_0 &=& |V-S-C|^2+ n_v\, \left[ |O-S|^2+|O-C|^2\right]\nn\\
 &&~~~~~~~~~~~~~~~~~+ (n_h - 1) \left[ (O-S)(\bar O - \bar C )+(O-C) (\bar O - \bar S)\right] +\ldots\nn\\
 &=& {\bf   G}_{2} + n_v {\bf   V}_{2} + n_h \, {\bf   H}_{2} \ ,
 \eea
 with $n_h$ and $n_v$ the number of hyper- and vector-multiplets 
 respectively, and
  \bea
{\bf   G}_{2} + {\bf  H}_{2} &=&  |V-S-C|^2 \ ,
\nn\\
{\bf V}_{2} &=& |O-S|^2+|O-C|^2 \ ,
\nn\\
{\bf  H}_{2} &=& (O-S)(\bar O - \bar C )+(O-C) (\bar O -
\bar S)
 \eea
  the ${\cal N}=2$ supergravity, hyper- and vector-multiplet contents,
  each comprising $4_B+4_F$ physical degrees of freedom.
  Due to the asymmetric twists and shifts, the resulting vacuum configurations
  do not correspond to
  compactifications of Type IIB on geometric CY manifolds, yet the theory enjoys $\cN =2$ spacetime susy.
  We are thus led to define the ``effective'' Hodge numbers
  \be
  h_{11} = n_h -1 \ , \quad \quad h_{21} = n_v \ ,
  \ee
and also define the ``effective'' Euler characteristic $\chi =
2(h_{11} - h_{21}) = 2(n_h-n_v) - 2$.

 In the following, we first describe in some details the simplest model with
 minimal massless content\footnote{A  related but different model with
extended $\cN=2_L + 2_R$ susy and thus larger massless multiplets
has been exhibited in \cite{Kiritsis:2008mu}.}, namely the one with  $(h_{11},h_{12})=(1,1)$.  Then, we
report the complete list of models resulting from our scan.

\subsection{An example: $(h_{11},h_{12})=(1,1)$ }

 One of the possible choices or twists and shifts that give rise to an interesting $(h_{11},h_{12})=(1,1)$ is the following:
 \bea
 b_1&=&  I _{3456} ~  \sigma_{1} ~    \overline{\sigma}_{5}   \ ,   \nn  \\
b_2 &=& I _{1256}    ~  \sigma_{3} ~  \overline{\sigma}_{12345}    \ ,     \nn  \\
  \bar{b}_1 &=& \bar{I}_{3456} ~ \sigma_{5}   ~    \overline{\sigma}_{1}   \ ,    \nn  \\
 \bar{b}_2 &=&  \bar{I }_{1256} ~ \sigma_{12345}    ~  \overline{\sigma}_{3}  \ .
\label{mod11}\eea
Many of the amplitudes vanish due to the presence of $SO(12)$ fermions in the odd spin structure.
The lattice sums of the non-vanishing amplitudes read
\bea
\begin{array}{llllll}
&&{\bf \L}[^{00}_{00}]=\ft12 \left( |\vartheta _{2}|^{12} + |\vartheta _{3}|^{12}+|\vartheta_{4}|^{12}\right)\nn\\
&&{\bf \L}[^{00}_{h0}]=\ft12 \vartheta _{3}^3 \vartheta _{4}^3 \bar{\vartheta }_{3} \bar{\vartheta }_{4} \left(\bar{\vartheta}_{3}^4+\bar{\vartheta }_{4}^4\right)    \nn\\
&&{\bf \L}[^{00}_{30}]=\ft12 |\vartheta _{3} \vartheta _{4}|^4 \left(\vartheta_{4}^2 \bar{\vartheta }_{3}^2+\vartheta _{3}^2 \bar{\vartheta }_{4}^2\right)\nn\\
&&{\bf \L}[^{00}_{h h'}]={\bf \L}[^{00}_{h 3}]= |\vartheta _{3} \vartheta _{4}|^6   \nn\\
&&{\bf \L}[^{00}_{33}]=\ft12  |\vartheta _{3} \vartheta _{4}|^4 \left(|\vartheta_{3}|^2+|\vartheta _{4}|^2\right)\nn\\
&&{\bf \L}[^{h 0}_{00}]=\ft12 \vartheta _{2}^3 \vartheta _{3}^3 \bar{\vartheta }_{2} \bar{\vartheta }_{3} \left(\bar{\vartheta}_{2}^4+\bar{\vartheta }_{3}^4\right)  \nn\\
&&{\bf \L}[^{30}_{00}]=\ft12 |\vartheta _{2} \vartheta _{3}|^4 \left(\vartheta_{3}^2 \bar{\vartheta }_{2}^2+\vartheta _{2}^2 \bar{\vartheta }_{3}^2\right) \nn\\
&&{\bf \L}[^{h h'}_{00}]={\bf \L}[^{h 3}_{00}]=  |\vartheta _{2} \vartheta _{3}|^6    \nn\\
&&{\bf \L}[^{33}_{00}]= \ft12 |\vartheta _{2} \vartheta _{3}|^4 \left(
|\vartheta_{2}|^4 + |\vartheta _{3}|^4 \right)\nn\\
&&{\bf \L}[^{h0}_{h0}]=\vartheta _{2}^3 \vartheta _{4}^3 \bar{\vartheta }_{2} \bar{\vartheta }_{4} \left(\bar{\vartheta}_{2}^4-\bar{\vartheta}_{4}^4\right)     \nn\\
&&{\bf \L}[^{30}_{30}]= \ft12 |\vartheta _{2} \vartheta _{4}|^4 \left(\vartheta_{2}^2 \bar{\vartheta }^2_{4}-\vartheta _{4}^2 \bar{\vartheta }_{2}^2\right) \nn\\
&&{\bf \L}[^{hh}_{h' h'}]= |\vartheta _{2} \vartheta _{4} |^6    \nn\\
&&{\bf \L}[^{33}_{33}]=\ft12 |\vartheta _{2} \vartheta _{4}|^4 \left(| \vartheta_{2} |^4+ | \vartheta _{4}|^4\right)
\end{array}
\eea
with $h,h'=1,2$ and ${\bf \L}[^{ab}_{cd}] = {\bf \L}[^{ba}_{dc}]^*$. Thanks
to the four independent $Z_2$ chiral twists all massless states in
the untwisted sector, except the $\cN = 2$ supergravity multiplet
and the universal dilaton hypermultiplet, are projected out. In
addition, due to the chiral shifts, most twisted sectors, except
for the $(3,3)$ sector, contribute only massive states. Indeed, a
twisted sector contribute massless states only when the shifts are
along the reflection plane. This condition is satisfied only for
$b_3 \bar b_3=I_{1234}\bar I_{1234} \sigma_{24}\bar
\sigma_{24}$.

As a consequence, the only massless contributions are: \bea
&&\sum_{a,b=0}^3 \cT_0[^{00}_{ab}] = |V-S-C|^2  \ , \nn\\
&&\cT_0[^{3  3}_{0 0}] + \cT_0[^{3  3}_{3  3}]=
|2 O-S-C|^2 \ . \eea
The latter, as anticipated, gives precisely one hyper- and one
vector- multiplet. This is the minimal massless content among all
known Type II compactifications admitting an isomorphism under
exchange of Left- and Right-movers and thus amenable to a natural
unoriented projection. We remark that several Left-Right
asymmetric models are known with fewer moduli:  for instance, a ``minimal''
$\cN=2_L + 0_R$ model with only the dilaton vector multiplet
recently exhibited in \cite{Dolivet:2007sz} as a starting point
for the construction of ``magic'' $\cN=2$ supergravity theories
\cite{MAGIC}. Moreover, there are models with $\cN=3$ susy
constructed in \cite{Ferrara:1989nm} with only one vector
multiplet, comprising only 3 complex massless scalars, including
the dilaton. Other systematic searches of models with low
``effective'' Hodge numbers \cite{Donagi:2008xy} seem to only
focus on Left-Right symmetric twists and shifts that lead at most
(or at least) to $h_{11} = h_{21} =3$, well known from the work of
Vafa and Witten \cite{Vafa:1994rv} and more recently of
\cite{Camara:2007dy} in the realm of Type I/heterotic duality.

\subsection{Various $\cN=1_L+1_R$ Models }

In addition to the above model, we have found many (new) Type IIB
non-geometric yet Left-Right symmetric models with low ``effective''
Hodge numbers, which may still turn out to be interesting starting
points for Type I model building.
We remark that although our models are given in terms of a rational
 CFT, a systematic study of open string descendants  is complicated
 by the high number of characters involved, typically of the order
 of one thousand. They can probably
 be explored by computer means along the lines of \cite{Kiritsis:2008mu}.

In Table $1$ we report all our consistent models.
 We keep track of the pattern of (pseudo)symmetry breaking
$SO(12)\rightarrow \prod_I SO(n_I)$.
Curiously, the
whole list of models can be grouped into the following three finite series for $(h_{11},h_{12})$:
\bea
(n,n) && \quad\quad n=1,2,3,4,5,9 \nn\\
(2n,2n+6), (2n+6,2n) &&  \quad\quad  n=0,1,2 \nn\\
(2n+3,2n+15), (2n+15,2n+3) && \quad\quad   n=0,1  \label{models}
\eea Our systematic search scans over all possible shifts with all
discrete torsion signs taken to be plus.  A longer list of
consistent models can be built after playing with more general
discrete torsion choices. In particular, it should be noticed that
the ``effective'' Euler number is always a multiple of 12, as
claimed in the Introduction. Finally, as apparent from Table $1$,
identical patterns of (pseudo)symmetry breaking may lead to rather
different massless spectra. This can be explained, in the cases
under consideration, by noticing that models with the same
breaking differ due to different choices of discrete torsions.

 The results of our systematic search partly overlap with the recent
results of a scan over $\Z_2$ orbifolds of the product of 18 Ising
models presented in \cite{Kiritsis:2008mu}. An important
difference between the two searches is  the choice of the $T^6$
lattice. We start with the non-factorizable $SO(12)$ maximal torus
$T^6_{SO(12)}$ while the authors of \cite{Kiritsis:2008mu} start
with a factorizable $T^6$.

\begin{center}
{\bf Table 1}
 \bea
\begin{array}{|c|c|c|}
\hline
~~~~~~~~~~ b's ~~~~~~~~~~&
~~~~~~~~~~~~~~SO(12)~~~~~~~~~~~~~~
& ~~~~~~~(h_{11},h_{12})~~~~~~~\\
%
%
\hline
\hline
\begin{array}{lllllll}
I _{3456}  ~   \sigma_{1} ~   \overline{\sigma}_{5}      \\
I _{1256}   ~   \sigma_{3} ~  \overline{\sigma}_{12345}      \\
\bar{I}_{3456} ~     \sigma_{5}  ~    \overline{\sigma}_{1}      \\
\bar{I }_{1256} ~     \sigma_{12345} ~  \overline{\sigma}_{3}
\end{array}
&
~~SO(2)^4\times O(1)^4 %
&
(1,1)\\
%
%
\hline
\begin{array}{lllllll}
I _{3456}  ~   \sigma_{1} ~    \overline{\sigma}_{2}       \\
I _{1256}   ~   \sigma_{3} ~   \overline{\sigma}_{12345}     \\
 \bar{I}_{3456} ~     \sigma_{2}  ~  \overline{\sigma}_{1}     \\
 \bar{I }_{1256} ~    \sigma_{12345}  ~   \overline{\sigma}_{3}
\end{array}
&
~~SO(3)\times SO(2)^2\times O(1)^5 &
(2,2)\\
%
%
\hline
\begin{array}{lllllll}
I _{3456}  ~   \sigma_{12} ~     \overline{\sigma}_{123456}       \\
I _{1256}   ~   \sigma_{236} ~   \overline{\sigma}_{1}     \\
\bar{I}_{3456} ~     \sigma_{123456} ~    \overline{\sigma}_{12}     \\
\bar{I }_{1256} ~     \sigma_{1}   ~   \overline{\sigma}_{236}
\end{array}
&
~~SO(3)^2\times SO(2)^2 \times O(1)^2
 &
(3,3)\\
%
%
\hline
\begin{array}{lllllll}
 I _{3456} ~   \sigma_{1} ~    \overline{\sigma}_{5}      \\
I _{1256}  ~   \sigma_{3} ~    \overline{\sigma}_{12456}       \\
\bar{I}_{3456} ~     \sigma_{5}  ~    \overline{\sigma}_{1}      \\
\bar{I }_{1256} ~     \sigma_{12456}  ~    \overline{\sigma}_{3}
\end{array}
&
~~SO(3)\times SO(2)^2\times O(1)^5 &
(4,4)\\
%
%
\hline
\begin{array}{lllllll}
 I _{3456}   ~  \sigma_{126} ~   \overline{\sigma}_{12}      \\
 I _{1256}  ~  \sigma_{346} ~   \overline{\sigma}_{35}      \\
  \bar{I }_{3456} ~    \sigma_{12}  ~  \overline{\sigma}_{126}      \\
  \bar{I }_{1256} ~    \sigma_{35}  ~  \overline{\sigma}_{346}
\end{array}
&
~~SO(2)^4\times O(1)^4
 &
(5,5)\\
%
%
\hline
\begin{array}{lllllll}
I_{3456} ~ \sigma_{12} ~\bar{\sigma }_{12} \\
I_{1256} ~ \sigma _{34} ~\bar{\sigma }_{56} \\
\overline{I}_{3456}~ \sigma _{12} ~ \bar{\sigma }_{12} \\
\overline{I}_{1256} ~  \sigma _{56}~ \bar{\sigma }_{34}
\end{array}
&
~~SO(2)^6 &
(9,9)\\
%
%
%
%
\hline
\begin{array}{lllllll}
 I_{3456} ~ \sigma_{12} ~
\bar{\sigma}_{13} \\
 I_{1256} ~ \sigma_{34} ~
\bar{\sigma}_{25} \\
\overline{I}_{3456} ~ \sigma_{13} ~  \bar{\sigma}_{12} \\
\overline{I}_{1256} ~ \sigma_{25} ~ \bar{\sigma}_{34}
\end{array}
&
~~SO(2)^3\times O(1)^6
&
(6,0)\\
%
%
%
%
\hline
\begin{array}{lllllll}
 I_{3456} ~ \sigma_{12} ~
\bar{\sigma}_{15} \\
 I_{1256} ~ \sigma_{34} ~
\bar{\sigma}_{36} \\
\overline{I}_{3456} ~\sigma_{15} ~
\bar{\sigma}_{12} \\
 \overline{I}_{1256} ~ \sigma_{36} ~
\bar{\sigma }_{34}
\end{array}
&
~~SO(2)^3\times O(1)^6
&
(0,6)\\
\hline
\end{array}
\nn
\eea

 \bea
\begin{array}{|c|c|c|}
\hline
~~~~~~~~~~ b's ~~~~~~~~~~&
~~~~~~~~~~~~~~SO(12)~~~~~~~~~~~~~~
& ~~~~~~~(h_{11},h_{12})~~~~~~~\\
%
%
%
\hline
\hline
\begin{array}{lllllll}
I_{3456} ~ \sigma_{1} ~
\bar{\sigma}_{4} \\
 I_{1256} ~ \sigma_{356} ~
\bar{\sigma}_{2} \\
\overline{I}_{3456} ~ \sigma_{4} ~
\bar{\sigma}_{1} \\
 \overline{I}_{1256} ~ \sigma_{2} ~
\bar{\sigma}_{356}
\end{array}
&
~~SO(3)^2 \times SO(2)\times O(1)^4
&
(2,8)\\
%
%
%
\hline
\begin{array}{lllllll}
 I_{3456} ~ \sigma_{1} ~
\bar{\sigma}_{2} \\
I_{1256} ~ \sigma_{356} ~
\bar{\sigma}_{4} \\
\overline{I}_{3456} ~ \sigma _{2} ~
\bar{\sigma}_{1} \\
\overline{I}_{1256} ~ \sigma_{4} ~
\bar{\sigma}_{356}
\end{array}
&
~~SO(3)^2 \times SO(2)\times O(1)^4
&
(8,2)\\
%
%
%
%
\hline
\begin{array}{lllllll}
 I_{3456} ~ \sigma_{1} ~
\bar{\sigma}_{5} \\
I_{1256} ~ \sigma_{346} ~
\bar{\sigma}_{25} \\
\overline{I}_{3456}~ \sigma_{5} ~
\bar{\sigma}_{1} \\
 \overline{I}_{1256} ~\sigma_{25} ~
\bar{\sigma}_{346}
\end{array}
&
~~SO(3)^2 \times SO(2)\times O(1)^4
&
(4,10)\\
%
%
%
%
\hline
\begin{array}{lllllll}
 I_{3456} ~ \sigma_{12} ~
\bar{\sigma}_{45} \\
 I_{1256} ~ \sigma_{36} ~
\bar{\sigma}_{5} \\
 \overline{I}_{3456} ~\sigma_{45} ~
\bar{\sigma}_{12} \\
\overline{I}_{1256} ~ \sigma_{5} ~
\bar{\sigma}_{36}
\end{array}
& ~~SO(3)^2 \times SO(2)\times O(1)^4
&
(10,4)\\
%
%
%
%
\hline
\begin{array}{lllllll}
  I _{3456}  ~   \sigma_{12} ~    \overline{\sigma}_{12}      \\
   I _{1256}  ~  \sigma_{34} ~   \overline{\sigma}_{34}       \\
 \bar{I }_{3456}  ~       \sigma_{12}  ~ \overline{\sigma}_{12}       \\
  \bar{I }_{1256}  ~      \sigma_{34}  ~ \overline{\sigma}_{34}
\end{array}
&
~~SO(2)^6 &
(15,3)\\
%
%
%
%
\hline
\begin{array}{lllllll}
I_{3456}~
\bar{\sigma}_{3456} \\
 I_{1256} ~
\bar{\sigma}_{1256} \\
 \overline{I}_{3456} ~ \sigma_{3456} \\
  \overline{I}_{1256} ~ \sigma_{1256}
\end{array}
&
~~SO(2)^6
&
(3,15)\\
%
%
%
%
\hline
\begin{array}{lllllll}
 I_{3456} ~ \sigma_{12} ~
\bar{\sigma}_{34} \\
 I_{1256} ~ \sigma_{34} ~
\bar{\sigma}_{123456} \\
\overline{I}_{3456} ~ \sigma_{34} ~
\bar{\sigma}_{12} \\
 \overline{I}_{1256} ~ \sigma_{123456} ~
\bar{\sigma}_{34}
 \end{array}
&
~~SO(4)\times SO(2)^4
&
(5,17)\\
%
%
%
\hline
\begin{array}{lllllll}
 I_{3456} ~ \sigma_{126} ~
\bar{\sigma}_{123456} \\
 I_{1256} ~\sigma_{5}~
\bar{\sigma}_{3456} \\
\overline{I}_{3456} ~ \sigma_{123456} ~
 \bar{\sigma}_{12} \\
 \overline{I}_{1256} ~ \sigma_{3456}~\bar \sigma_{5}
\end{array}
&
~~SO(4)\times SO(2)^4
&
(17,5)\\
\hline

\end{array}
\nn
\eea

 \bea
\begin{array}{|c|c|c|}
\hline
~~~~~~~~~~ b's ~~~~~~~~~~&
~~~~~~~~~~~~~~SO(12)~~~~~~~~~~~~~~
& ~~~~~~~(h_{11},h_{12})~~~~~~~\\

%
%
\hline
\hline
\begin{array}{lllllll}
 I_{3456} ~ \sigma_{1} ~
\bar{\sigma}_{12456} \\
 I_{1256} ~\sigma_{356}~
\bar{\sigma}_{23456} \\
\overline{I}_{3456} ~ \sigma_{12456} ~
 \bar{\sigma}_{1} \\
 \overline{I}_{1256} ~ \sigma_{23456}~\bar \sigma_{356}
\end{array}
&
~~SO(3)^2\times SO(2)\times O(1)^4
&
(6,12)\\
\hline
%
%
\hline
\begin{array}{lllllll}
 I_{3456} ~ \sigma_{1} ~
\bar{\sigma}_{23456} \\
 I_{1256} ~\sigma_{356}~
\bar{\sigma}_{12456} \\
\overline{I}_{3456} ~ \sigma_{23456} ~
 \bar{\sigma}_{1} \\
 \overline{I}_{1256} ~ \sigma_{12456}~\bar \sigma_{356}
\end{array}
&
~~SO(3)^2\times SO(2)\times O(1)^4
&
(12,6)\\
\hline
\end{array}
\nn\eea
\end{center}

\section{Models with $\cN=1_L $}

Another class of interesting Type II models are the Left-Right
asymmetric orbifolds with $\cN=1_L $ spacetime susy. In the
bosonic description, these models arise from including a
projection $(-)^{F_R}\sigma_R$, thus breaking all supersymmetries
associated to the Right-movers and preventing any of those to
reappear in the twisted sectors by means of the order two
chiral shift $\sigma_R$. In the fermionic description, $\sigma_R$ simply amount to 
a reflection of all the $SO(12)$ fermions.  Thus, the projection is equivalent to 
choose a basis of sets consisting only of $F$ and $S$.
This is the starting point of our systematic
search in this largely unexplored class of Type II vacuum
configurations.
The resulting $\cN =4_L + 0_R $ spectrum is
coded in the one-loop torus amplitude (\ref{torus2}).
Supersymmetric massless states only arise from the combination
$(V_8 - S_8) O_{12} \bar V_{20}$, that produces $\cN=4$
supergravity coupled to 18 vector multiplets. A careful look at
the corresponding vertex operators and their OPE's shows that the
gauge group is $SU(2)^6$, as a remnant of the structure of the
internal world-sheet cubic supercurrent \cite{ABKW}. This or an
equivalent model has been found in the seminal paper
\cite{Dixon:1987yp}. The emergence of Right-moving world-sheet
currents, generating a supersymmetric Kac-Moody algebra, has been
deeply analyzed  in view of the possibility of producing
non-abelian NS gauge symmetries. The authors of
\cite{Dixon:1987yp}
 arrived however at the negative conclusion that (perturbative)
Type II models cannot accommodate the Standard Model with its
matter content.

To the sets $F$ and $S$ we have added two more sets $b_1$ and
$b_2$ producing a breaking of spacetime susy down to $\cN=1_L +
0_R $ and, at the same time, a breaking of the internal
(pseudo)symmetry $SO(20)$. Indeed, what we said in the context of
the above $\cN =4_L+ 0_R $ model applies to $\cN =1_L+ 0_R $, too.
The ``true'' gauge symmetry can only be determined after a careful
analysis of the vertex operators for the vector fields and their
OPE's, while taking into account the precise structure of the
cubic supercurrent. Since the only cubic supercurrent we consider
is expressed in terms of the $SU(2)^6$ structure constants, the
resulting gauge symmetry we find is a subgroup of $SU(2)^6$ with
abelian factors. Moreover, there are massless charged chiral
multiplets that can further break the gauge symmetry by a
perturbative  Higgs mechanism.

In the following we describe in some details a specific model with
 minimal number of chiral multiplets and then collect the remaining
models in table $2$. The massless spectrum decomposes according to
\bea \cT_0={\bf G}_{1}   +n_v{\bf V}_{1}  +n_{v'}{\bf V}'_{1} +n_c
{\bf C}_{1} + n_{c'}  {\bf C}'_{1} \ , \eea with \bea
{\bf G}_{1}   + {\bf C}_{1}  &=& (V-S-C) \,\bar V  \ , \nn\\
 {\bf V}_{1} &=& (V-S-C) \,\bar O \nn\\
 {\bf V}'_{1} &=&  S \bar S+C \bar C-S\bar O-C\bar O  \ , \nn\\
{\bf C}_{1} &=& (2O-S-C) \,\bar O
  \nn\\
{\bf C}'_{1} &=&  C\bar S+S\bar C-O\bar S-O\bar C
  \eea the content of
the gravity, vector and chiral  multiplets, and $n_v+n_{v'}$,
$n_c+ n_{c'}$   the total numbers of vector and chiral multiplets. Although
primed and unprimed multiplets have identical field content, we
find it convenient to distinguish them in order to stress the
different origin, NS-NS or R-R, of their bosonic degrees of
freedom. It is amusing to stress that generalized D-branes
\cite{Bianchi:2008cj} and their exotic open string excitations can
be introduced that couple to the twisted R-R states.

\subsection{An example: $(n_v,n_v';n_c,n_c ')=(14, 0;5, 0)$   }

Let us discuss the  model with  generators
 \bea
\begin{array}{lllllll}
b_{1}&=&  I _{3456} ~ \sigma_{12} ~   \overline{\sigma}_{45}  \ ,  \\
b_{2}&=&  I _{1256} ~ \sigma_{36} ~  \overline{\sigma}_{5}  \ .
\end{array}
\eea
In addition to breaking spacetime supersymmetry to $\cN =1$, the
two $Z_2$ actions break the internal (pseudo)symmetry according to
\be SO(12)_L\times SO(20)_R\to \left[ SO(4)^2\times
SO(2)^2\right]_L\times \left[  SO(2)^2\times SO(16)\right] _R  \ .
\ee Actually, $SO(16)_R \rightarrow SO(2)\times SO(14)$, where the
first factor is the little group for massless particles in $D=4$.

 The non-vanishing lattice sums read \bea
&&{\bf \L}[^0_0]=
\ft 12 \left(\vartheta _{3}^6 \bar{\vartheta }_{3}^{10}-\vartheta _{4}^6
\bar{\vartheta }_{4}^{10}-\vartheta _{2}^6 \bar{\vartheta }_{2}^{10}\right)
\nn\\
&&{\bf \L}[^0_1]=\ft 12
\vartheta _{3}^2 \vartheta _{4}^2 \bar{\vartheta }_{3}^2 \bar{\vartheta }_{4}^2 \left(\vartheta
   _{4}^2 \bar{\vartheta }_{3}^6-\vartheta _{3}^2 \bar{\vartheta}_{4}^6\right)\nn\\
&&{\bf \L}[^1_0]=\ft 12
\vartheta _{2}^2 \vartheta _{3}^2  \bar{\vartheta }_{2}^2 \bar{\vartheta }_{3}^2 \left(\vartheta
   _{2}^2 \bar{\vartheta }_{3}^6-\vartheta _{3}^2 \bar{\vartheta }_{2}^6\right)
\nn\\
&&{\bf \L}[^1_1]=\ft 12
\vartheta _{2}^2 \vartheta _{4}^2 \bar{\vartheta }_{2}^2 \bar{\vartheta }_{4}^2 \left(\vartheta
   _{2}^2 \bar{\vartheta }_{4}^6-\vartheta _{4}^2\bar{\vartheta }_{2}^6\right)
\nn\\
&&{\bf \L}[^0_2]={\bf \L}[^0_3]=\ft 12
\vartheta _{3}^3 \vartheta _{4}^3 \bar{\vartheta }_{3} \bar{\vartheta }_{4} \left(\bar{\vartheta }_{3}^8- \bar{\vartheta }_{4}^8\right)
\nn\\
&&{\bf \L}[^2_0]={\bf \L}[^3_0]=\ft 12
\vartheta _{2}^3 \vartheta _{3}^3 \bar{\vartheta }_{2} \bar{\vartheta }_{3} \left( \bar{\vartheta }_{3}^8-\bar{\vartheta }_{2}^8\right)
\nn\\
&&{\bf \L}[^2_2]={\bf \L}[^3_3]=\ft 12
\vartheta _{2}^3 \vartheta _{4}^3 \bar{\vartheta }_{2} \bar{\vartheta }_{4} \left(\bar{\vartheta }_{4}^8- \bar{\vartheta }_{2}^8\right)
 \ .
\eea
Massless states come only from the untwisted sector leading to
\bea \cT_0=(V-S-C) (\bar V + 14 \bar O) + 4(2O-S-C) \bar O = {\bf
G}_{1}  + 14 {\bf V}_{1} + 5 {\bf C}_{1}    \ . \eea The resulting
gauge group is $SU(2)^4\times U(1)^2$. The universal chiral
multiplet is neutral, while the additional four chiral multiplets
are charged with respect to the abelian factors. They form two
pairs of charge $(\pm 1, 0)$ and $(0,\pm 1)$. Along the flat
directions of the D-term potential, the $U(1)^2$ gauge symmetry is
generically broken. Since no matter fields are charged with
respect to $SU(2)^6$, the latter remains as an unbroken gauge
symmetry in perturbation theory. It would be very important to
study the possibility of including both physical and Euclidean
Left-Right asymmetric D-branes  in the background in order to have
a richer matter spectrum and turn on non-perturbative effects.

\subsection{Various $\cN=1_L$ Models }

Table $2$ summarizes the results of our preliminary search of Type
IIB models with $\cN = 1_L + 0_R$. The basis sets are now, besides
the universal $F$ and $S$,  the two additional \bea &&
b_1 = I_{3456}\, \sigma^{i_1 i_2 \ldots }\,\bar \sigma^{k_1 k_2 \ldots } = \{(\psi\, y)^{3456}  \, (y\, w)^{i_1 i_2 \ldots }  | (\tilde y\, \tilde w)^{k_1 k_2 \ldots }  \}  \ , \nn\\
&&b_2 = I_{1256}\, \sigma^{j_1 j_2 \ldots }\,\bar \sigma^{l_1 l_2 \ldots } = \{(\psi\, y)^{1256}  \, (y\, w)^{j_1 j_2 \ldots }  | (\tilde y\, \tilde w)^{ l_1 l_2 \ldots }   \}  \ ,  \eea
  with the scanning that runs over all choices of the sets  $ (i_1  i_2 \ldots) $,  $(j_1 j_2 \ldots) $,   $ (k_1  k_2 \ldots) $,  $(l_1 l_2 \ldots) $, compatibly again with the conditions (\ref{consistency}).
  Each set $b_\a$ breaks half of the spacetime susy's arising from
the Left-moving sector, while supersymmetry associated to the
Right-moving sectors is completely broken to start with. As
apparent from Table 2, the reduction in the number of moduli in
$\cN= 1_L$ models is less significant than in $\cN= 1_L + 1_R$
models. This is due to the presence of the tachyonic vacuum in the
R-moving sector that, combined with internal excitations, can
produce physical (\ie level matched) particle states.
\begin{center}
{\bf Table 2}
\bea
\begin{array}{|c|c|c|}
\hline
~~~~~~~~~~ b's ~~~~~~~~~~&
~~~~~~~~SO(12)_{L}\times SO(20)_R~~~~~~~~
& ~~~~~~~(n_v,n_{v'} ; n_c,n_{c'} )~~~~~~~\\
%
%
%
\hline
\hline
\begin{array}{lllll}
 I_{3456} ~ \sigma_{12} ~
\bar{\sigma}_{45} \\
 I_{1256} ~ \sigma_{36} ~
\bar{\sigma}_{5}
\end{array}
& ~~ \left[SO(4)^2\times SO(2)^2\right]_L\times \left[SO(16)\times
SO(2)^2\right]_R
&
(14,0;5,0)\\
\hline

\begin{array}{lllll}
 I_{3456} ~ \sigma_{126} ~
\bar{\sigma}_{12} \\
 I_{1256} ~ \sigma_{346} ~
\bar{\sigma}_{35}
\end{array}
& ~~ \left[SO(6)\times SO(2)^3\right]_L\times \left[SO(4)^2\times
SO(12)\right]_R
&
(10,0;25,0)\\
\hline

\begin{array}{lllllll}
 I_{3456} ~ \sigma_{1} ~
\bar{\sigma}_{5} \\
 I_{1256} ~ \sigma_{3} ~
\bar{\sigma}_{12345}
\end{array}
&
~~ \left[SO(4)^2\times SO(2)^2\right]_L\times \left[SO(8)\times SO(2)\times SO(10)\right]_R %
&
(8,0;27,0)\\
\hline

\begin{array}{lllll}
 I_{3456} ~ \sigma_{12} ~
\bar{\sigma}_{123456} \\
 I_{1256} ~ \sigma_{236} ~
\bar{\sigma}_{1}
\end{array}
& ~~ \left[SO(4)^2\times SO(2)^2\right]_L\times \left[SO(2)\times
SO(10)\times SO(8)\right]_R
&
(6,8;13,8)\\
\hline

%
\begin{array}{lllll}
 I_{3456} ~ \sigma_{12} ~
\bar{\sigma}_{34} \\
 I_{1256} ~\sigma_{34} ~
\bar{\sigma}_{123456}
\end{array}
&
~~ \left[SO(6)\times SO(2)^3\right]_L\times \left[SO(4)\times SO(8)\times SO(8)\right]_R %
&
(6,8;29,8)\\
\hline
%
%
\end{array}
\nn\eea
\end{center}

\section{Unoriented projections}

The Left-Right symmetric Type IIB string vacua we constructed in
section \ref{sect11} admit a natural $\Omega$ projection. It is an
interesting question whether they can be taken as a starting point
of orientifold constructions with phenomenologically interesting
open string chiral matter.
For unoriented strings \cite{BPS}, several closed string moduli
are odd under
  $\Omega$ and are thus projected out. Hypermultiplets of the oriented $\cN=2$ theory
  reduce to $\cN=1$ chiral multiplets while vector multiplets lead to vector or chiral multiplets
  according to their parity under $\Omega$. In addition,
  D-brane sectors should be added
  in the presence of non-trivial closed string tadpoles.

  Here we discuss  the simplest instances
of unoriented projections with and without open strings.

\subsection{The minimal model}

 We start by considering the unoriented projection of the $\cN=1_L+1_R$ model
with $(h_{11}= h_{21})=(1,1)$ discussed in Section 4.1,
corresponding to the choice of generators in eq. (\ref{mod11}).
Notice that, since $\Omega$ identifies Left and Right movers, one has
\be
    {\rm Tr}_{\cH_L\otimes \cH_R } \,
    \Omega \,(g^L \otimes g^R)=  {\rm Tr}_{\cH_L} \, g_{_\Omega}  \ ,
     \ee
where  $g_{_\Omega}$ is the diagonal action  $g^L  g^R$ with Left
and Right moving fields identified, i.e. $\bar I_i \to I_i$, $\bar
\sigma_i \to \sigma_i$ . In this way, the $g_{_\Omega}$ amplitudes
are not the naive chiral halves of the amplitudes entering the
torus amplitude.  They must be rather written in terms of traces
over the chiral modes  of the $g_\Omega$ orbifold group generators
corresponding to the sets \bea && b_{1\Omega}= I_{3456}
\sigma_{15} =\{   \psi^{3456} \, y^{1346}\, w^{15} \} \ , \nn\\
&&b_{2\Omega}= I_{1256} ~\sigma_{1245}=\{   \psi^{1256} \, y^{46}\, w^{1245} \}  \ , \nn\\
&&b_{3\Omega}= I_{1234}  ~\sigma_{24}=\{   \psi^{1234} \, y^{13}\, w^{24} \} \ , \eea
where $b_{3\Omega}=b_{1\Omega} \ b_{2\Omega}$.
In addition, only Left-Right symetrically twisted states enter the Klein-bottle amplitude.
 In the direct channel one then gets
 \bea \cK &=& \frac{1}{16} \sum_{a,b,c,d} {\rm Tr}_{\cH_{Lc}\otimes \cH_{Rd} } \Omega\, b_{a}\, \bar b_b
 = \frac{1}{4} \sum_{a,b}{\rm Tr}_{\cH_{La}}  b_{b\Omega}= \frac{1}{4 \ \eta^8}  \sum_{a,b=0}^3
\epsilon_{a,b} \, \rho_{ab} \, {\Lambda}[^a_{b}] \
\eea
where the unoriented lattice sums read
\bea &&{\Lambda}[^0_{0}] = \vartheta_3^6 +\epsilon\, \vartheta_2^6 \\
&&{\Lambda}[^0_{h}]  = \vartheta_4^3 \vartheta_3^3 +\epsilon\, \vartheta_1^3\vartheta_2^3 \\
&&{\Lambda}[^0_{3}]  = \vartheta_4^2 \vartheta_3^4 +\epsilon\, \vartheta_1^2\vartheta_2^4 \\
&&{\Lambda}[^3_{0}]  = \vartheta_2^2 \vartheta_3^4 +\epsilon\, \vartheta_3^2\vartheta_2^4 \\
&&{\Lambda}[^3_{3}]  = \vartheta_1^2 \vartheta_3^4 +\epsilon\, \vartheta_4^2\vartheta_2^4 \\
&&{\Lambda}[^h_{0}] = \vartheta_2^3 \vartheta_3^3 +\epsilon\, \vartheta_3^3\vartheta_2^3 \\
&&{\Lambda}[^h_{h}] = \vartheta_1^3 \vartheta_3^3 + \epsilon\,\vartheta_4^3\vartheta_2^3  \ .
\eea
$\epsilon_{a,b}$ and $\epsilon$ are signs satisfying the fusion
constraints and $h=1,2$.  All  the other possible lattice sums
vanish. For instance,
\be {\Lambda}[^3_{h}] = \vartheta_3\vartheta_4^3
\vartheta_1^2\vartheta_2^2 +
\vartheta_2\vartheta_1^3\vartheta_3^2\vartheta_4^2 \equiv 0  \ .
\ee

Performing an $S$ modular transformation one can determine the
Klein-bottle amplitude in the transverse channel
 \be \tilde{\cK} = \frac{2^2}{4 \ \eta^8}  \sum_{a,b=0}^3
\epsilon_{b,a} \,\sigma_{b,a} \,  \rho_{ab} \, {\tilde\Lambda}[^a_{b}]\label{ktra}  \ , \ee
where
 \bea &&{\tilde\Lambda}[^0_{0}]  = \vartheta_3^6 +\epsilon\, \vartheta_4^6 \\
&&{\tilde\Lambda}[^0_{h}]  = \vartheta_4^3 \vartheta_3^3 + \epsilon\,\vartheta_3^3\vartheta_4^3 \\
&&{\tilde\Lambda}[^0_{3}]  = \vartheta_4^2 \vartheta_3^4 + \epsilon\, \vartheta_3^2\vartheta_4^4 \\
&&{\tilde\Lambda}[^3_{0}]  = \vartheta_2^2 \vartheta_3^4 + \epsilon\, \vartheta_1^2\vartheta_4^4 \\
&&{\tilde\Lambda}[^3_{3}]  = \vartheta_1^2 \vartheta_3^4 + \epsilon\, \vartheta_2^2\vartheta_4^4 \\
&&{\tilde\Lambda}[^h_{0}] = \vartheta_2^3 \vartheta_3^3 + \epsilon\, \vartheta_1^3\vartheta_4^3 \\
&&{\tilde\Lambda}[^h_{h}] = \vartheta_1^3 \vartheta_3^3 + \epsilon\, \vartheta_2^3\vartheta_4^3  \ .
\eea
with $\sigma_{ab}$  some signs given in (\ref{sphases}).
Choosing  $\epsilon=-1$ and all the remaining signs $\epsilon_{a,b}=1$,
one finds that no massless untwisted or twisted tadpoles are present.
The unoriented model is then consistent by itself and no D-branes are needed.
At the massless level one finds
\be
\cK_{\rm massless}=(V-S-C)+(2O-S-C)
\ee
Together with the torus contribution one is left with the minimal $\cN=1$ content
\be
 \ft12(\cT+\cK)_{\rm massless}=  {\bf G}_{1}+  2\, {\bf C}_{1}  \ .
\ee

\subsection{Models with open strings}

Here we present the simplest instance
of an unoriented projection with open strings. For simplicity we consider the case
of  $T^6/\Z_{2L}\times \Z_{2L}'\times\Z_{2R}\times\Z_{2R}'$ with
no shifts. As before, we take the $T^6$  at the $SO(12)$ point. The
orbifold group generators are \be b_1=I_{3456} \ , \quad
b_2=I_{1256} \ , \quad \bar b_1=\bar I_{3456}  \ , \quad \bar
b_2=\bar I_{1256}  \ .\ee The resulting model can be written in
terms of 64 characters collecting the chiral states in the
$a$-twisted sector ($a=0,1,2,3$) with $ \Z_{2L}\times \Z_{2L}$
eigenvalues $(\pm,\pm)$ in one of the four O, V, S, C conjugacy
classes of the $SO(12)$ lattice. The complete list of characters
can be found in Appendix B. In particular, orbifold group
invariant states in the untwisted sector are labelled by $\chi
_{1}, \chi_5,\chi_{9}, \chi_{13}$. The untwisted torus is then
given by \be {\cal T}_{\rm unt}=|\chi_{1}|^2+|
\chi_5|^2+|\chi_{9}|^2+| \chi_{13}|^2 \  . \label{tunt} \ee
  The twisted amplitudes complete (\ref{tunt})
  in a modular invariant form with positive integer coefficients.
  We discuss the two possibilities
  \bea
  {\cal T}_A&=& |\chi_{1}+\chi_{17}+\chi_{35}+\chi_{49}|^2+|\chi_{5}+\chi_{21}+\chi_{39}+\chi_{53}|^2\nn\\
&&
+ |\chi_{9}+\chi_{30}+\chi_{45}+\chi_{64}|^2
+|\chi_{13}+\chi_{26}+\chi_{41}+\chi_{60}|^2  \ , \\
{\cal T}_B &=&
{\chi}_{1} \, {\overline{\chi  }}_{1}+{\chi  }_{18} \, {\overline{\chi  }}_{2}+{\chi  }_{33} \, {\overline{\chi  }}_{3}+{\chi  }_{52} \, {\overline{\chi  }}_{4}+{\chi  }_{5} \, {\overline{\chi  }}_{5}+{\chi  }_{22} \, {\overline{\chi  }}_{6}+{\chi  }_{37} \, {\overline{\chi  }}_{7}+{\chi  }_{56} \, {\overline{\chi  }}_{8}+{\chi  }_{9} \, {\overline{\chi  }}_{9}\nn\\
&&
+{\chi  }_{29} \, {\overline{\chi  }}_{10}+{\chi  }_{47} \, {\overline{\chi  }}_{11}+{\chi  }_{61} \, {\overline{\chi  }}_{12}+{\chi  }_{13} \, {\overline{\chi  }}_{13}+{\chi  }_{25} \, {\overline{\chi  }}_{14}+{\chi  }_{43} \, {\overline{\chi  }}_{15}+{\chi  }_{57} \, {\overline{\chi  }}_{16}+{\chi  }_{17} \, {\overline{\chi  }}_{17}\nn\\
&&+{\chi  }_{2} \, {\overline{\chi  }}_{18}+{\chi  }_{51} \, {\overline{\chi  }}_{19}+{\chi  }_{34} \, {\overline{\chi  }}_{20}+{\chi  }_{21} \, {\overline{\chi  }}_{21}+{\chi  }_{6} \, {\overline{\chi  }}_{22}+{\chi  }_{55} \, {\overline{\chi  }}_{23}+{\chi  }_{38} \, {\overline{\chi  }}_{24}+{\chi  }_{14} \, {\overline{\chi  }}_{25}\nn\\
&&+{\chi  }_{26} \, {\overline{\chi  }}_{26}+{\chi  }_{44} \, {\overline{\chi  }}_{27}+{\chi  }_{58} \, {\overline{\chi  }}_{28}+{\chi  }_{10} \, {\overline{\chi  }}_{29}+{\chi  }_{30} \, {\overline{\chi  }}_{30}+{\chi  }_{48} \, {\overline{\chi  }}_{31}+{\chi  }_{62} \, {\overline{\chi  }}_{32}+{\chi  }_{3} \, {\overline{\chi  }}_{33}\nn\\
&&+{\chi  }_{20} \, {\overline{\chi  }}_{34}+{\chi  }_{35} \,
{\overline{\chi  }}_{35}+{\chi  }_{50} \, {\overline{\chi
}}_{36}+{\chi  }_{7} \, {\overline{\chi  }}_{37}+{\chi  }_{24} \,
{\overline{\chi  }}_{38}+{\chi  }_{39} \, {\overline{\chi
}}_{39}+{\chi  }_{54} \, {\overline{\chi  }}_{40}+{\chi  }_{41} \,
{\overline{\chi  }}_{41} \nn\\&&+{\chi  }_{59} \, {\overline{\chi
}}_{42}+{\chi  }_{15} \, {\overline{\chi  }}_{43}+{\chi  }_{27} \,
{\overline{\chi  }}_{44}+{\chi  }_{45} \, {\overline{\chi
}}_{45}+{\chi  }_{63} \, {\overline{\chi  }}_{46}+{\chi  }_{11} \,
{\overline{\chi  }}_{47}+{\chi  }_{31} \, {\overline{\chi
}}_{48}+{\chi  }_{49} \, {\overline{\chi  }}_{49} \nn\\&&+{\chi
}_{36} \, {\overline{\chi  }}_{50}+{\chi  }_{19} \,
{\overline{\chi  }}_{51}+{\chi  }_{4} \, {\overline{\chi
}}_{52}+{\chi  }_{53} \, {\overline{\chi  }}_{53}+{\chi  }_{40} \,
{\overline{\chi  }}_{54}+{\chi  }_{23} \, {\overline{\chi
}}_{55}+{\chi  }_{8} \, {\overline{\chi  }}_{56}+{\chi  }_{16} \,
{\overline{\chi  }}_{57} \nn\\&&+{\chi  }_{28} \, {\overline{\chi
}}_{58}+{\chi  }_{42} \, {\overline{\chi  }}_{59}+{\chi  }_{60} \,
{\overline{\chi  }}_{60}+ {\chi  }_{12} \, {\overline{\chi
}}_{61}+{\chi  }_{32} \, {\overline{\chi  }}_{62}+{\chi  }_{46} \,
{\overline{\chi  }}_{63}+{\chi  }_{64} \, {\overline{\chi  }}_{64}
\  .
  \eea
  They coincide in the untwisted sector and are distinguished by the
  pairing of states in the twisted sectors, namely they
   correspond to different choices of discrete torsion giving rise to different modular invariants
   \cite{Bianchi:1999uq}.
   In particular, case ${\bf A}$ corresponds to a modular invariant
   with extended symmetry that gives back the
  toroidal compactification
  of Type IIB on the $T^6$ based on the lattice of $SO(12)$.
  On the other hand, case ${\bf B}$ corresponds to a permutation
  modular invariant with effective Hodge numbers $(15,15)$. Indeed,
  out of the 64 original characters, the set of massless characters
  consists in
  \be
\{ \chi_{1}, {\chi  }_{2}, {\chi  }_{3},{\chi  }_{4},{\chi  }_{17},{\chi  }_{18},{\chi  }_{23},{\chi  }_{24},{\chi  }_{33},{\chi  }_{35},{\chi  }_{38},{\chi  }_{40},{\chi  }_{49},{\chi  }_{52},{\chi  }_{54},{\chi  }_{55} \} \ ,
\ee
   with ${\chi  }_{1}=V-S-C+\ldots $ and
   $\chi_{i}=2O-S-C +\ldots $ for the remaining ones.
   Plugging the above expansions into the expressions
   for the two torus amplitudes one finds
   \bea
  ({\cal T}_{A})_{\rm massless}&=& |V+6O-4S-4C|^2 \ , \nn\\
  ({\cal T}_{B})_{\rm massless}&=& |V-S-C|^2+15|2O-S-C|^2 \ .
     \eea
   The Klein-bottle amplitude
   follows from ${\cal T}_A$ and $ {\cal T}_B$ by reducing to
   their diagonal components. In both cases
   one finds
   \bea
   {\cal K}&=& \chi_{1}+\chi_{17}+\chi_{35}+\chi_{49}+\chi_{5}+\chi_{21}+\chi_{39}+\chi_{53}\nn\\
&& + \chi_{9}+\chi_{30}+\chi_{45}+\chi_{64}+\chi_{13}
+\chi_{26}+\chi_{41}+\chi_{60} \ ,
   \eea
   that produces
       \bea
   {\cal K}&=& (V-S-C)+3(2O-S-C)
      \eea
   at the massless level.  The unoriented projection results in case ${\bf A}$ into the supegravity multiplet with 6 vector multiplets of ${\cal N}=4$, while in case ${\bf B}$
   it leads to  ${\cal N}=1$ supergravity with 6 vector multiplets and
   25 chiral multiplets.

Going to the transverse channel one finds
 \be
  \tilde {\cal K}= 2^3(\chi_{1}+\chi_{17}+\chi_{35}+\chi_{49} ) \ .
   \ee
 The tadpoles can be cancelled by adding
 the tranverse Annulus and Moebius amplitudes
 \bea
 \tilde {\cal A} &=& 2^{-3} (\chi_{1}+\chi_{17}+\chi_{35}+\chi_{49} )(n_1+n_2+\bar n_1+\bar n_2)^2\nn\\
 &&+ 2^{-3}(\chi_{5}+\chi_{21}+\chi_{39}+\chi_{53})(n_1-n_2+\bar n_1-\bar n_2)^2 \nn\\
&& + 2^{-3}(\chi_{9}+\chi_{30}+\chi_{45}+\chi_{64})(n_1+n_2-\bar n_1-\bar n_2)^2 \nn\\
&&+2^{-3}(\chi_{13}+\chi_{26}+\chi_{41}+\chi_{60})(n_1-n_2-\bar n_1+\bar n_2)^2 \ , \\
 \tilde {\cal M} &=&  - (\chi_{1}+\chi_{17}+\chi_{35}+\chi_{49} )(n_1+n_2+\bar n_1+\bar n_2) \ ,
 \eea
 provided
 \be
 n_1+n_2=4 \ .
 \ee

 Finally, applying $S$ and $P=T^{1\over 2} S T^2 S T^{1\over 2}$ modular transformations one
 finds the direct amplitudes
  \bea
 {\cal A} &=&  (\chi_{1}+\chi_{17}+\chi_{35}+\chi_{49} )
 (2 n_1\bar n_1+ 2 n_2 \bar n_2)\nn\\
 && + (\chi_{5}+\chi_{21}+\chi_{39}+\chi_{53})
 (n_1^2+n_2^2+\bar n_1^2+\bar n_2^2) \nn\\
&& + (\chi_{9}+\chi_{30}+\chi_{45}+\chi_{64})
(2 n_1 n_2+ 2\bar n_1 \bar n_2) \nn\\
&&+(\chi_{13}+\chi_{26}+\chi_{41}+\chi_{60})
(2 n_1 \bar n_2+ 2 n_2 \bar n_1) \ , \\
 {\cal M} &=&  (\chi_{5}+\chi_{21}+\chi_{39}+\chi_{53} )
(n_1+n_2+\bar n_1+\bar n_2) \ .
 \eea
 The massless open string spectrum, encoded in $( \cA + \cM)/2$, is that of
 ${\cal N}=4$ SYM with gauge group $U(N)\times U(4-N)$. Notice that in case
 ${\bf B}$ only an ${\cal N}=1$ fraction of the ${\cal N}=4$
 brane supersymmetry is preserved by the bulk theory. An analogous
 behavior can be observed in other cases, most notably the open
 descendants of the
 $D_{odd}$ series of $SU(2)$ WZW models \cite{Pradisi:1995pp, Pradisi:1996yd}

 \section{Conclusions and perspectives}

In perturbative string theory, moduli fields are exactly marginal
deformations of the underlying conformal field theory. In the low
energy description, they correspond to perturbatively exact flat
directions of the scalar potential.
In the present paper, we have exploited  $\Z_2$ chiral twists and
shifts in the search of calculable Type IIB models with few
moduli. We have explored both Left-Right symmetric, though
non-geometric, models with $\cN = 1_L + 1_R$ spacetime susy and
Left-Right asymmetric models with $\cN = 1_L + 0_R$ spacetime
susy. We have found a finite series of models enjoying $\cN = 1_L +
1_R$ spacetime susy with very low ``effective'' Hodge numbers
$(h_{11},h_{21})$ given by \bea
(n,n) && \quad\quad n=1,2,3,4,5,9 \nn\\
(2n,2n+6), (2n+6,2n) &&  \quad\quad  n=0,1,2 \nn\\
(2n+3,2n+15), (2n+15,2n+3) && \quad\quad   n=0,1 \eea Most of
these models have no counterpart in previous CY or RCFT scans
\cite{Kiritsis:2008mu, Donagi:2008xy}. We have studied the
``minimal'' model with $h_{11}=h_{21}=1$  in details and
constructed one of its $\cN=1$ unoriented descendants with no open
strings. This model exhibits the minimal (as far as we know)
$\cN=1$ field content found so far in the moduli space of
perturbative string compactifications. We cannot exclude the
possibility that more general  chiral twists and shifts could give
rise to perturbative Type IIB models with $\cN =1$ spacetime susy
and only the universal dilaton
 chiral multiplet or to an $\cN =2$ model
 with $h_{11}=h_{21}=0$\footnote{A  Left-Right asymmetric ``minimal'' model with $\cN =2_L + 0_R$
spacetime susy and only the dilaton vector (!) multiplet has been
constructed by similar means in \cite{Dolivet:2007sz}
but does not admit an obvious
unoriented projection.}.

Our main motivation was  to identify convenient starting points
for calculable orientifold constructions exhibiting complete
moduli stabilization. We find that asymmetric twists and shifts
can be easily combined in order to freeze out most closed string
moduli. The effect on open string moduli is subtler. The only
model with open unoriented strings, we have analyzed in some
detail, enjoys extended $\cN=4$ susy in the open sector and is
thus non-chiral. Apparently there is some tension between
chirality and moduli stabilization\footnote{P.~Camara and others
share our viewpoint.}. The interesting question of whether
D-branes with phenomenologically viable gauge group and chiral
matter contents can be accommodated in this picture remains open.

\section*{Acknowledgments}

 We would like to thank C.~Angelantonj, E.~Dudas,
S.~Ferrara, E.~Kiritsis, C.~Kounnas, K.~Narain, A.~Sagnotti, B.~
Schellekens, Ya.~Stanev, and M.~C.~Timirgaziu for interesting
discussions. Preliminary results were presented by M.~B. at  {\it
Vacuum Selection in String Theory} (Liverpool University, March
2008), at {\it String Phenomenology '08} (U. Penn, Philadelphia,
May-June 2008), at  {\it Pre-Strings Phenomenology} (CERN, July
2008), at the $4^{th}$ RTN Meeting {\it Symmetries and Structure
of the Universe} (Varna, September 2008), and at {\it Mathematical
Challenges in String Phenomenology} (ESI, Vienna, October 2008).
M.~B. would like to thank the organizers for creating a
stimulating environment and the participants for making useful
comments. This work was supported in part by the MIUR-PRIN
contract 2007-5ATT78, NATO PST.CLG.978785, the RTN grants MRTNCT-
2004-503369, EU MRTN-CT-2004-512194, and MRTN-CT-2004-005104.
P.~A. would like to thank the Physics Departments at University of
Rome ``Tor Vergata'' and at University of Crete in Heraklion for
hospitality during completion of this work.

\appendix
\section{Some definitions}
%

In this Appendix we collect some useful formulas illustrating our conventions.
We adopt the following definition for the Jacobi theta functions:
\bea
 \vartheta[^a_b](v|\t)&=&\sum_{n\in \Zint} q^{{1\over 2}(n+a)^2}
e^{2\p i (n+a)(v+b)} \ .
\eea
The Characters of $SO(2n)$ level one are
\bea O_{2n} &=&{1\over
2\eta^n}(\vartheta_3^n+\vartheta_4^n) \quad ;\quad
V_{2n} ={1\over 2\eta^n}(\vartheta_3^n-\vartheta_4^n) \ ; \nn\\
S_{2n} &=&{1\over 2\eta^n}(\vartheta_2^n+i^{-n} \vartheta_1^n)
\quad ;\quad C_{2n} ={1\over
2\eta^n}(\vartheta_2^n-i^{-n}\vartheta_1^n) \ ,
 \eea
and the corresponding ground states can be described as
 \bea
 q^{1\over 6}\, O_{4} &=& (1,1)+\ldots  \ ,  \nn\\
  q^{1\over 6}\, V_{4} &=& (2,2) \, q^{1\over 2}+\ldots  \ , \nn\\
   q^{1\over 6}\, S_{4} &=& (1,2)\, q^{1\over 4}+\ldots  \ , \nn\\
    q^{1\over 6}\, C_{4} &=& (2,1)\, q^{1\over 4}+\ldots  \ .\nn\\
 \eea

The modular transformation matrices on the characters of $SO(2n)$ level one are the following
   \bea
 T &=& e^{-{i\p n \over 12}} \ {\rm Diag} \, (1,-1,e^{i\pi n\over 4},   e^{i\pi n\over 4} )  \ , \nn\\~\\
S &=&{1\over 2}\left(\ba{rrrr} 1&1&1&1\\
                      1&1&-1&-1\\
                      1&-1&i^{-n}&-i^{-n}\\
                      1&-1&-i^{-n}&i^{-n}  \ea\right) \ , \nn\\~\\
P &=&     \left(\ba{rrrr} c&s&0&0\\
                      s&-c&0&0\\
                      0&0& c\, \xi& i s\, \xi\\
                      0&0& i s \,\xi & c \,\xi  \ea\right)  \ ,
\eea with $s=\sin {n\pi\over 4}$, $c=\cos {n\pi\over 4}$ and
$\xi=e^{-{i n\pi\over 4}}$.

The space-time characters for the supersymmetric $Z_2 \times Z_2$
model are \cite{MBthesis}
  \bea
\t_{00} &=& V_2O_2O_2O_2 + O_2V_2V_2V_2 - S_2S_2S_2S_2 - C_2C_2C_2C_2\nn\\
\t_{01} &=& O_2V_2O_2O_2 + V_2O_2V_2V_2 - C_2C_2S_2S_2 - S_2S_2C_2C_2\nn\\
\t_{02} &=& O_2O_2V_2O_2 + V_2V_2O_2V_2 - C_2S_2C_2S_2 - S_2C_2S_2C_2\nn\\
\t_{03} &=& O_2O_2O_2V_2 + V_2V_2V_2O_2 - C_2S_2S_2C_2 - S_2C_2C_2S_2\nn\\
\t_{10} &=& V_2O_2S_2C_2 + O_2V_2C_2S_2 - S_2S_2V_2O_2 - C_2C_2O_2V_2\nn\\
\t_{11} &=& O_2V_2S_2C_2 + V_2O_2C_2S_2 - S_2S_2O_2V_2 - C_2C_2V_2O_2\nn\\
\t_{12} &=& O_2O_2C_2C_2 + V_2V_2S_2S_2 - S_2C_2V_2V_2 - C_2S_2O_2O_2\nn\\
\t_{13} &=& O_2O_2S_2S_2 + V_2V_2C_2C_2 - C_2S_2V_2V_2 - S_2C_2O_2O_2\nn\\
\t_{20} &=& V_2S_2O_2C_2 + O_2C_2V_2S_2 - S_2V_2S_2O_2 - C_2O_2C_2V_2\nn\\
\t_{21} &=& O_2C_2O_2C_2 + V_2S_2V_2S_2 - C_2O_2S_2O_2 - S_2V_2C_2V_2\nn\\
\t_{22} &=& O_2S_2V_2C_2 + V_2C_2O_2S_2 - C_2V_2C_2O_2 - S_2O_2S_2V_2\nn\\
\t_{23} &=& O_2S_2O_2S_2 + V_2C_2V_2C_2 - C_2V_2S_2V_2 - S_2O_2C_2O_2\nn\\
\t_{30} &=& V_2S_2C_2O_2 + O_2C_2S_2V_2 - C_2O_2V_2C_2 - S_2V_2O_2S_2\nn\\
\t_{31} &=& O_2C_2C_2O_2 + V_2S_2S_2V_2 - C_2O_2O_2S_2 - S_2V_2V_2C_2\nn\\
\t_{32} &=& O_2S_2S_2O_2 + V_2C_2C_2V_2 - C_2V_2V_2S_2 - S_2O_2O_2C_2\nn\\
\t_{33} &=& O_2S_2C_2V_2 + V_2C_2S_2O_2 - S_2O_2V_2S_2 -
C_2V_2O_2C_2  \ ,  \eea
while the corresponding amplitudes $\rho_{a,b}$ expressed in terms of the previous characters result: \bea
  \rho_{a0}   &=&  \tau_{a0}+\tau_{a1}+\tau_{a2}+\tau_{a3} \ , \nn\\
  \rho_{a1}    &=&  \tau_{a0}+\tau_{a1}-\tau_{a2}-\tau_{a3} \ , \nn\\
  \rho_{a2}   &=&   \tau_{a0}-\tau_{a1}+\tau_{a2}-\tau_{a3} \ , \nn\\
  \rho_{a3}   &=&  \tau_{a0}-\tau_{a1}-\tau_{a2}+\tau_{a3} \ .
\label{tou's}  \eea

It is useful to recall their $S$-modular transformations

\be
\rho_{ab}(-1/{\tau}) \ = \ \sigma_{ab} \ \rho_{ba}(\tau)  \ ,
\ee
where the phases are
\bea
&\sigma_{00}& \ = \ \sigma_{01} \ = \ \sigma_{02} \ = \ \sigma_{03} \ = \ \sigma_{10} \ = \ \sigma_{20} \ = \ \sigma_{30} \ = \ 1  \quad ,\nn\\
&\sigma_{11}& \ = \ \sigma_{22} \ = \ \sigma_{33} \ = \ -1  \quad ,\nn\\
&\sigma_{13}& \ = \ - \sigma_{12} \ = \ - \sigma_{21} \ = \ - \sigma_{23} \ = \ \sigma_{31} \ = \ - \sigma_{32} \ = \ i  \quad . \label{sphases}
\eea

\section{Characters of $T^6/\Z_{2L}\times\Z_{2L}'\times \Z_{2R}\times\Z_{2R}'$ }

In this Appendix we list the 64 characters corresponding to the
chiral amplitudes of the Type IIB compactification on
$T^6/\Z_{2L}\times\Z_{2L}'\times \Z_{2R}\times\Z_{2R}'$ that enter
the partition functions of the models discussed in Section 6.
{\tiny{
\bea
\chi_{1}&=&
\left(O_2 O_2 O_2 O_6+V_2 V_2
   V_2 V_6\right) \tau _{00}+\left(O_2 V_2 V_2
   O_6+V_2 O_2 O_2 V_6\right) \tau
   _{01}+\left(V_2 O_2 V_2 O_6+O_2 V_2
   O_2 V_6\right) \tau _{02}+\left(V_2 V_2 O_2
   O_6+O_2 O_2 V_2 V_6\right) \tau
   _{03}\nn\\  \chi_{2}&=&        \left(O_2 V_2 V_2 O_6+V_2 O_2
   O_2 V_6\right) \tau _{00}+\left(O_2 O_2 O_2
   O_6+V_2 V_2 V_2 V_6\right) \tau
   _{01}\   +\left(V_2 V_2 O_2 O_6+O_2 O_2
   V_2 V_6\right) \tau _{02}+\left(V_2 O_2 V_2
   O_6+O_2 V_2 O_2 V_6\right) \tau
   _{03}\nn\\  \chi_{3}&=&        \left(V_2 O_2 V_2 O_6+O_2 V_2
   O_2 V_6\right) \tau _{00}+\left(V_2 V_2 O_2
   O_6+O_2 O_2 V_2 V_6\right) \tau
   _{01}\   +\left(O_2 O_2 O_2 O_6+V_2 V_2
   V_2 V_6\right) \tau _{02}+\left(O_2 V_2 V_2
   O_6+V_2 O_2 O_2 V_6\right) \tau
   _{03}\nn\\  \chi_{4}&=&        \left(V_2 V_2 O_2 O_6+O_2 O_2
   V_2 V_6\right) \tau _{00}+\left(V_2 O_2 V_2
   O_6+O_2 V_2 O_2 V_6\right) \tau
   _{01}\   +\left(O_2 V_2 V_2 O_6+V_2 O_2
   O_2 V_6\right) \tau _{02}+\left(O_2 O_2 O_2
   O_6+V_2 V_2 V_2 V_6\right) \tau
   _{03}\nn\\  \chi_{5}&=&        \left(V_2 V_2 V_2 O_6+O_2 O_2
   O_2 V_6\right) \tau _{00}+\left(V_2 O_2 O_2
   O_6+O_2 V_2 V_2 V_6\right) \tau
   _{01}\   +\left(O_2 V_2 O_2 O_6+V_2 O_2
   V_2 V_6\right) \tau _{02}+\left(O_2 O_2 V_2
   O_6+V_2 V_2 O_2 V_6\right) \tau
   _{03}\nn\\  \chi_{6}&=&        \left(V_2 O_2 O_2 O_6+O_2 V_2
   V_2 V_6\right) \tau _{00}+\left(V_2 V_2 V_2
   O_6+O_2 O_2 O_2 V_6\right) \tau
   _{01}\   +\left(O_2 O_2 V_2 O_6+V_2 V_2
   O_2 V_6\right) \tau _{02}+\left(O_2 V_2 O_2
   O_6+V_2 O_2 V_2 V_6\right) \tau
   _{03}\nn\\  \chi_{7}&=&        \left(O_2 V_2 O_2 O_6+V_2 O_2
   V_2 V_6\right) \tau _{00}+\left(O_2 O_2 V_2
   O_6+V_2 V_2 O_2 V_6\right) \tau
   _{01}\   +\left(V_2 V_2 V_2 O_6+O_2 O_2
   O_2 V_6\right) \tau _{02}+\left(V_2 O_2 O_2
   O_6+O_2 V_2 V_2 V_6\right) \tau
   _{03}\nn\\  \chi_{8}&=&        \left(O_2 O_2 V_2 O_6+V_2 V_2
   O_2 V_6\right) \tau _{00}+\left(O_2 V_2 O_2
   O_6+V_2 O_2 V_2 V_6\right) \tau
   _{01}\   +\left(V_2 O_2 O_2 O_6+O_2 V_2
   V_2 V_6\right) \tau _{02}+\left(V_2 V_2 V_2
   O_6+O_2 O_2 O_2 V_6\right) \tau
   _{03}\nn\\  \chi_{9}&=&        \left(C_2 C_2 C_2 C_6+S_2 S_2
   S_2 S_6\right) \tau _{00}+\left(C_2 S_2 S_2
   C_6+S_2 C_2 C_2 S_6\right) \tau
   _{01}\   +\left(S_2 C_2 S_2 C_6+C_2 S_2
   C_2 S_6\right) \tau _{02}+\left(S_2 S_2 C_2
   C_6+C_2 C_2 S_2 S_6\right) \tau
   _{03}\nn\\  \chi_{10}&=&        \left(C_2 S_2 S_2 C_6+S_2 C_2
   C_2 S_6\right) \tau _{00}+\left(C_2 C_2 C_2
   C_6+S_2 S_2 S_2 S_6\right) \tau
   _{01}\   +\left(S_2 S_2 C_2 C_6+C_2 C_2
   S_2 S_6\right) \tau _{02}+\left(S_2 C_2 S_2
   C_6+C_2 S_2 C_2 S_6\right) \tau
   _{03}\nn\\  \chi_{11}&=&        \left(S_2 C_2 S_2 C_6+C_2 S_2
   C_2 S_6\right) \tau _{00}+\left(S_2 S_2 C_2
   C_6+C_2 C_2 S_2 S_6\right) \tau
   _{01}\   +\left(C_2 C_2 C_2 C_6+S_2 S_2
   S_2 S_6\right) \tau _{02}+\left(C_2 S_2 S_2
   C_6+S_2 C_2 C_2 S_6\right) \tau
   _{03}\nn\\  \chi_{12}&=&        \left(S_2 S_2 C_2 C_6+C_2 C_2
   S_2 S_6\right) \tau _{00}+\left(S_2 C_2 S_2
   C_6+C_2 S_2 C_2 S_6\right) \tau
   _{01}\   +\left(C_2 S_2 S_2 C_6+S_2 C_2
   C_2 S_6\right) \tau _{02}+\left(C_2 C_2 C_2
   C_6+S_2 S_2 S_2 S_6\right) \tau
   _{03}\nn\\  \chi_{13}&=&        \left(S_2 S_2 S_2 C_6+C_2 C_2
   C_2 S_6\right) \tau _{00}+\left(S_2 C_2 C_2
   C_6+C_2 S_2 S_2 S_6\right) \tau
   _{01}\   +\left(C_2 S_2 C_2 C_6+S_2 C_2
   S_2 S_6\right) \tau _{02}+\left(C_2 C_2 S_2
   C_6+S_2 S_2 C_2 S_6\right) \tau
   _{03}\nn\\  \chi_{14}&=&        \left(S_2 C_2 C_2 C_6+C_2 S_2
   S_2 S_6\right) \tau _{00}+\left(S_2 S_2 S_2
   C_6+C_2 C_2 C_2 S_6\right) \tau
   _{01}\   +\left(C_2 C_2 S_2 C_6+S_2 S_2
   C_2 S_6\right) \tau _{02}+\left(C_2 S_2 C_2
   C_6+S_2 C_2 S_2 S_6\right) \tau
   _{03}\nn\\  \chi_{15}&=&        \left(C_2 S_2 C_2 C_6+S_2 C_2
   S_2 S_6\right) \tau _{00}+\left(C_2 C_2 S_2
   C_6+S_2 S_2 C_2 S_6\right) \tau
   _{01}\   +\left(S_2 S_2 S_2 C_6+C_2 C_2
   C_2 S_6\right) \tau _{02}+\left(S_2 C_2 C_2
   C_6+C_2 S_2 S_2 S_6\right) \tau
   _{03}\nn\\  \chi_{16}&=&        \left(C_2 C_2 S_2 C_6+S_2 S_2
   C_2 S_6\right) \tau _{00}+\left(C_2 S_2 C_2
   C_6+S_2 C_2 S_2 S_6\right) \tau
   _{01}\   +\left(S_2 C_2 C_2 C_6+C_2 S_2
   S_2 S_6\right) \tau _{02}+\left(S_2 S_2 S_2
   C_6+C_2 C_2 C_2 S_6\right) \tau
   _{03}\nn\\  \chi_{17}&=&        \left(V_2 S_2 S_2 O_6+O_2
   C_2 C_2 V_6\right) \tau _{10}+\left(V_2 C_2
   C_2 O_6+O_2 S_2 S_2 V_6\right) \tau
   _{11}\   +\left(O_2 S_2 C_2 O_6+V_2 C_2
   S_2 V_6\right) \tau _{12}+\left(O_2 C_2 S_2
   O_6+V_2 S_2 C_2 V_6\right) \tau
   _{13}\nn\\  \chi_{18}&=&        \left(V_2 C_2 C_2 O_6+O_2 S_2
   S_2 V_6\right) \tau _{10}+\left(V_2 S_2 S_2
   O_6+O_2 C_2 C_2 V_6\right) \tau
   _{11}\   +\left(O_2 C_2 S_2 O_6+V_2 S_2
   C_2 V_6\right) \tau _{12}+\left(O_2 S_2 C_2
   O_6+V_2 C_2 S_2 V_6\right) \tau
   _{13}\nn\\  \chi_{19}&=&        \left(O_2 S_2 C_2 O_6+V_2 C_2
   S_2 V_6\right) \tau _{10}+\left(O_2 C_2 S_2
   O_6+V_2 S_2 C_2 V_6\right) \tau
   _{11}\   +\left(V_2 S_2 S_2 O_6+O_2 C_2
   C_2 V_6\right) \tau _{12}+\left(V_2 C_2 C_2
   O_6+O_2 S_2 S_2 V_6\right) \tau
   _{13}\nn\\  \chi_{20}&=&        \left(O_2 C_2 S_2 O_6+V_2 S_2
   C_2 V_6\right) \tau _{10}+\left(O_2 S_2 C_2
   O_6+V_2 C_2 S_2 V_6\right) \tau
   _{11}\   +\left(V_2 C_2 C_2 O_6+O_2 S_2
   S_2 V_6\right) \tau _{12}+\left(V_2 S_2 S_2
   O_6+O_2 C_2 C_2 V_6\right) \tau
   _{13} \nn\\  \chi_{21}&=&        \left(O_2 C_2 C_2 O_6+V_2
   S_2 S_2 V_6\right) \tau _{10}+\left(O_2 S_2
   S_2 O_6+V_2 C_2 C_2 V_6\right) \tau
   _{11}\   +\left(V_2 C_2 S_2 O_6+O_2 S_2
   C_2 V_6\right) \tau _{12}+\left(V_2 S_2 C_2
   O_6+O_2 C_2 S_2 V_6\right) \tau
   _{13}\nn\\  \chi_{22}&=&        \left(O_2 S_2 S_2 O_6+V_2 C_2
   C_2 V_6\right) \tau _{10}+\left(O_2 C_2 C_2
   O_6+V_2 S_2 S_2 V_6\right) \tau
   _{11}\   +\left(V_2 S_2 C_2 O_6+O_2 C_2
   S_2 V_6\right) \tau _{12}+\left(V_2 C_2 S_2
   O_6+O_2 S_2 C_2 V_6\right) \tau
   _{13}\nn\\  \chi_{23}&=&        \left(V_2 C_2 S_2 O_6+O_2 S_2
   C_2 V_6\right) \tau _{10}+\left(V_2 S_2 C_2
   O_6+O_2 C_2 S_2 V_6\right) \tau
   _{11}\   +\left(O_2 C_2 C_2 O_6+V_2 S_2
   S_2 V_6\right) \tau _{12}+\left(O_2 S_2 S_2
   O_6+V_2 C_2 C_2 V_6\right) \tau
   _{13}\nn\\  \chi_{24}&=&        \left(V_2 S_2 C_2 O_6+O_2 C_2
   S_2 V_6\right) \tau _{10}+\left(V_2 C_2 S_2
   O_6+O_2 S_2 C_2 V_6\right) \tau
   _{11}\   +\left(O_2 S_2 S_2 O_6+V_2 C_2
   C_2 V_6\right) \tau _{12}+\left(O_2 C_2 C_2
   O_6+V_2 S_2 S_2 V_6\right) \tau
   _{13}\nn\\  \chi_{25}&=&        \left(C_2 O_2 O_2 C_6+S_2
   V_2 V_2 S_6\right) \tau _{10}+\left(C_2 V_2
   V_2 C_6+S_2 O_2 O_2 S_6\right) \tau
   _{11}\   +\left(S_2 O_2 V_2 C_6+C_2 V_2
   O_2 S_6\right) \tau _{12}+\left(S_2 V_2 O_2
   C_6+C_2 O_2 V_2 S_6\right) \tau
   _{13}\nn\\  \chi_{26}&=&        \left(C_2 V_2 V_2 C_6+S_2 O_2
   O_2 S_6\right) \tau _{10}+\left(C_2 O_2 O_2
   C_6+S_2 V_2 V_2 S_6\right) \tau
   _{11}\   +\left(S_2 V_2 O_2 C_6+C_2 O_2
   V_2 S_6\right) \tau _{12}+\left(S_2 O_2 V_2
   C_6+C_2 V_2 O_2 S_6\right) \tau
   _{13}\nn\\  \chi_{27}&=&        \left(S_2 O_2 V_2 C_6+C_2 V_2
   O_2 S_6\right) \tau _{10}+\left(S_2 V_2 O_2
   C_6+C_2 O_2 V_2 S_6\right) \tau
   _{11}\   +\left(C_2 O_2 O_2 C_6+S_2 V_2
   V_2 S_6\right) \tau _{12}+\left(C_2 V_2 V_2
   C_6+S_2 O_2 O_2 S_6\right) \tau
   _{13}\nn\\  \chi_{28}&=&        \left(S_2 V_2 O_2 C_6+C_2 O_2
   V_2 S_6\right) \tau _{10}+\left(S_2 O_2 V_2
   C_6+C_2 V_2 O_2 S_6\right) \tau
   _{11}\   +\left(C_2 V_2 V_2 C_6+S_2 O_2
   O_2 S_6\right) \tau _{12}+\left(C_2 O_2 O_2
   C_6+S_2 V_2 V_2 S_6\right) \tau
   _{13}\nn\\  \chi_{29}&=&        \left(S_2 V_2 V_2 C_6+C_2
   O_2 O_2 S_6\right) \tau _{10}+\left(S_2 O_2
   O_2 C_6+C_2 V_2 V_2 S_6\right) \tau
   _{11}\   +\left(C_2 V_2 O_2 C_6+S_2 O_2
   V_2 S_6\right) \tau _{12}+\left(C_2 O_2 V_2
   C_6+S_2 V_2 O_2 S_6\right) \tau
   _{13}\nn\\  \chi_{30}&=&        \left(S_2 O_2 O_2 C_6+C_2 V_2
   V_2 S_6\right) \tau _{10}+\left(S_2 V_2 V_2
   C_6+C_2 O_2 O_2 S_6\right) \tau
   _{11}\   +\left(C_2 O_2 V_2 C_6+S_2 V_2
   O_2 S_6\right) \tau _{12}+\left(C_2 V_2 O_2
   C_6+S_2 O_2 V_2 S_6\right) \tau
   _{13}\nn\\  \chi_{31}&=&        \left(C_2 V_2 O_2 C_6+S_2 O_2
   V_2 S_6\right) \tau _{10}+\left(C_2 O_2 V_2
   C_6+S_2 V_2 O_2 S_6\right) \tau
   _{11}\   +\left(S_2 V_2 V_2 C_6+C_2 O_2
   O_2 S_6\right) \tau _{12}+\left(S_2 O_2 O_2
   C_6+C_2 V_2 V_2 S_6\right) \tau
   _{13}\nn\\  \chi_{32}&=&        \left(C_2 O_2 V_2 C_6+S_2 V_2
   O_2 S_6\right) \tau _{10}+\left(C_2 V_2 O_2
   C_6+S_2 O_2 V_2 S_6\right) \tau
   _{11}\   +\left(S_2 O_2 O_2 C_6+C_2 V_2
   V_2 S_6\right) \tau _{12}+\left(S_2 V_2 V_2
   C_6+C_2 O_2 O_2 S_6\right) \tau
   _{13}\nn\\  \chi_{33}&=&        \left(S_2 V_2 S_2 O_6+C_2
   O_2 C_2 V_6\right) \tau _{20}+\left(S_2 O_2
   C_2 O_6+C_2 V_2 S_2 V_6\right) \tau
   _{21}\   +\left(C_2 V_2 C_2 O_6+S_2 O_2
   S_2 V_6\right) \tau _{22}+\left(C_2 O_2 S_2
   O_6+S_2 V_2 C_2 V_6\right) \tau
   _{23}\nn\\  \chi_{34}&=&        \left(S_2 O_2 C_2 O_6+C_2 V_2
   S_2 V_6\right) \tau _{20}+\left(S_2 V_2 S_2
   O_6+C_2 O_2 C_2 V_6\right) \tau
   _{21}\   +\left(C_2 O_2 S_2 O_6+S_2 V_2
   C_2 V_6\right) \tau _{22}+\left(C_2 V_2 C_2
   O_6+S_2 O_2 S_2 V_6\right) \tau
   _{23}\nn\\  \chi_{35}&=&        \left(C_2 V_2 C_2 O_6+S_2 O_2
   S_2 V_6\right) \tau _{20}+\left(C_2 O_2 S_2
   O_6+S_2 V_2 C_2 V_6\right) \tau
   _{21}\   +\left(S_2 V_2 S_2 O_6+C_2 O_2
   C_2 V_6\right) \tau _{22}+\left(S_2 O_2 C_2
   O_6+C_2 V_2 S_2 V_6\right) \tau
   _{23}\nn\\  \chi_{36}&=&        \left(C_2 O_2 S_2 O_6+S_2 V_2
   C_2 V_6\right) \tau _{20}+\left(C_2 V_2 C_2
   O_6+S_2 O_2 S_2 V_6\right) \tau
   _{21}\   +\left(S_2 O_2 C_2 O_6+C_2 V_2
   S_2 V_6\right) \tau _{22}+\left(S_2 V_2 S_2
   O_6+C_2 O_2 C_2 V_6\right) \tau
   _{23}\nn\\  \chi_{37}&=&        \left(C_2 O_2 C_2 O_6+S_2
   V_2 S_2 V_6\right) \tau _{20}+\left(C_2 V_2
   S_2 O_6+S_2 O_2 C_2 V_6\right) \tau
   _{21}\   +\left(S_2 O_2 S_2 O_6+C_2 V_2
   C_2 V_6\right) \tau _{22}+\left(S_2 V_2 C_2
   O_6+C_2 O_2 S_2 V_6\right) \tau
   _{23}\nn\\  \chi_{38}&=&        \left(C_2 V_2 S_2 O_6+S_2 O_2
   C_2 V_6\right) \tau _{20}+\left(C_2 O_2 C_2
   O_6+S_2 V_2 S_2 V_6\right) \tau
   _{21}\   +\left(S_2 V_2 C_2 O_6+C_2 O_2
   S_2 V_6\right) \tau _{22}+\left(S_2 O_2 S_2
   O_6+C_2 V_2 C_2 V_6\right) \tau
   _{23}\nn\\  \chi_{39}&=&        \left(S_2 O_2 S_2 O_6+C_2 V_2
   C_2 V_6\right) \tau _{20}+\left(S_2 V_2 C_2
   O_6+C_2 O_2 S_2 V_6\right) \tau
   _{21}\   +\left(C_2 O_2 C_2 O_6+S_2 V_2
   S_2 V_6\right) \tau _{22}+\left(C_2 V_2 S_2
   O_6+S_2 O_2 C_2 V_6\right) \tau
   _{23}\nn\\  \chi_{40}&=&        \left(S_2 V_2 C_2 O_6+C_2 O_2
   S_2 V_6\right) \tau _{20}+\left(S_2 O_2 S_2
   O_6+C_2 V_2 C_2 V_6\right) \tau
   _{21}\   +\left(C_2 V_2 S_2 O_6+S_2 O_2
   C_2 V_6\right) \tau _{22}+\left(C_2 O_2 C_2
   O_6+S_2 V_2 S_2 V_6\right) \tau
   _{23}\nn\\  \chi_{41}&=&        \left(O_2 C_2 O_2 C_6+V_2
   S_2 V_2 S_6\right) \tau _{20}+\left(O_2 S_2
   V_2 C_6+V_2 C_2 O_2 S_6\right) \tau
   _{21}\   +\left(V_2 C_2 V_2 C_6+O_2 S_2
   O_2 S_6\right) \tau _{22}+\left(V_2 S_2 O_2
   C_6+O_2 C_2 V_2 S_6\right) \tau
   _{23}\nn\\  \chi_{42}&=&        \left(O_2 S_2 V_2 C_6+V_2 C_2
   O_2 S_6\right) \tau _{20}+\left(O_2 C_2 O_2
   C_6+V_2 S_2 V_2 S_6\right) \tau
   _{21}\   +\left(V_2 S_2 O_2 C_6+O_2 C_2
   V_2 S_6\right) \tau _{22}+\left(V_2 C_2 V_2
   C_6+O_2 S_2 O_2 S_6\right) \tau
   _{23}\nn\\  \chi_{43}&=&        \left(V_2 C_2 V_2 C_6+O_2 S_2
   O_2 S_6\right) \tau _{20}+\left(V_2 S_2 O_2
   C_6+O_2 C_2 V_2 S_6\right) \tau
   _{21}\   +\left(O_2 C_2 O_2 C_6+V_2 S_2
   V_2 S_6\right) \tau _{22}+\left(O_2 S_2 V_2
   C_6+V_2 C_2 O_2 S_6\right) \tau
   _{23}\nn\\  \chi_{44}&=&        \left(V_2 S_2 O_2 C_6+O_2 C_2
   V_2 S_6\right) \tau _{20}+\left(V_2 C_2 V_2
   C_6+O_2 S_2 O_2 S_6\right) \tau
   _{21}\   +\left(O_2 S_2 V_2 C_6+V_2 C_2
   O_2 S_6\right) \tau _{22}+\left(O_2 C_2 O_2
   C_6+V_2 S_2 V_2 S_6\right) \tau
   _{23}\nn\\  \chi_{45}&=&        \left(V_2 S_2 V_2 C_6+O_2
   C_2 O_2 S_6\right) \tau _{20}+\left(V_2 C_2
   O_2 C_6+O_2 S_2 V_2 S_6\right) \tau
   _{21}\   +\left(O_2 S_2 O_2 C_6+V_2 C_2
   V_2 S_6\right) \tau _{22}+\left(O_2 C_2 V_2
   C_6+V_2 S_2 O_2 S_6\right) \tau
   _{23}\nn\\  \chi_{46}&=&        \left(V_2 C_2 O_2 C_6+O_2 S_2
   V_2 S_6\right) \tau _{20}+\left(V_2 S_2 V_2
   C_6+O_2 C_2 O_2 S_6\right) \tau
   _{21}\   +\left(O_2 C_2 V_2 C_6+V_2 S_2
   O_2 S_6\right) \tau _{22}+\left(O_2 S_2 O_2
   C_6+V_2 C_2 V_2 S_6\right) \tau
   _{23}\nn\\  \chi_{47}&=&        \left(O_2 S_2 O_2 C_6+V_2 C_2
   V_2 S_6\right) \tau _{20}+\left(O_2 C_2 V_2
   C_6+V_2 S_2 O_2 S_6\right) \tau
   _{21}\   +\left(V_2 S_2 V_2 C_6+O_2 C_2
   O_2 S_6\right) \tau _{22}+\left(V_2 C_2 O_2
   C_6+O_2 S_2 V_2 S_6\right) \tau
   _{23}\nn\\  \chi_{48}&=&        \left(O_2 C_2 V_2 C_6+V_2 S_2
   O_2 S_6\right) \tau _{20}+\left(O_2 S_2 O_2
   C_6+V_2 C_2 V_2 S_6\right) \tau
   _{21}\   +\left(V_2 C_2 O_2 C_6+O_2 S_2
   V_2 S_6\right) \tau _{22}+\left(V_2 S_2 V_2
   C_6+O_2 C_2 O_2 S_6\right) \tau
   _{23}\nn\\  \chi_{49}&=&        \left(S_2 S_2 V_2 O_6+C_2
   C_2 O_2 V_6\right) \tau _{30}+\left(S_2 C_2
   O_2 O_6+C_2 S_2 V_2 V_6\right) \tau
   _{31}\   +\left(C_2 S_2 O_2 O_6+S_2 C_2
   V_2 V_6\right) \tau _{32}+\left(C_2 C_2 V_2
   O_6+S_2 S_2 O_2 V_6\right) \tau
   _{33}\nn\\  \chi_{50}&=&        \left(S_2 C_2 O_2 O_6+C_2 S_2
   V_2 V_6\right) \tau _{30}+\left(S_2 S_2 V_2
   O_6+C_2 C_2 O_2 V_6\right) \tau
   _{31}\   +\left(C_2 C_2 V_2 O_6+S_2 S_2
   O_2 V_6\right) \tau _{32}+\left(C_2 S_2 O_2
   O_6+S_2 C_2 V_2 V_6\right) \tau
   _{33}\nn\\  \chi_{51}&=&        \left(C_2 S_2 O_2 O_6+S_2 C_2
   V_2 V_6\right) \tau _{30}+\left(C_2 C_2 V_2
   O_6+S_2 S_2 O_2 V_6\right) \tau
   _{31}\   +\left(S_2 S_2 V_2 O_6+C_2 C_2
   O_2 V_6\right) \tau _{32}+\left(S_2 C_2 O_2
   O_6+C_2 S_2 V_2 V_6\right) \tau
   _{33}\nn\\  \chi_{52}&=&        \left(C_2 C_2 V_2 O_6+S_2 S_2
   O_2 V_6\right) \tau _{30}+\left(C_2 S_2 O_2
   O_6+S_2 C_2 V_2 V_6\right) \tau
   _{31}\   +\left(S_2 C_2 O_2 O_6+C_2 S_2
   V_2 V_6\right) \tau _{32}+\left(S_2 S_2 V_2
   O_6+C_2 C_2 O_2 V_6\right) \tau
   _{33}\nn\\  \chi_{53}&=&        \left(C_2 C_2 O_2 O_6+S_2
   S_2 V_2 V_6\right) \tau _{30}+\left(C_2 S_2
   V_2 O_6+S_2 C_2 O_2 V_6\right) \tau
   _{31}\   +\left(S_2 C_2 V_2 O_6+C_2 S_2
   O_2 V_6\right) \tau _{32}+\left(S_2 S_2 O_2
   O_6+C_2 C_2 V_2 V_6\right) \tau
   _{33}\nn\\  \chi_{54}&=&        \left(C_2 S_2 V_2 O_6+S_2 C_2
   O_2 V_6\right) \tau _{30}+\left(C_2 C_2 O_2
   O_6+S_2 S_2 V_2 V_6\right) \tau
   _{31}\   +\left(S_2 S_2 O_2 O_6+C_2 C_2
   V_2 V_6\right) \tau _{32}+\left(S_2 C_2 V_2
   O_6+C_2 S_2 O_2 V_6\right) \tau
   _{33}\nn\\  \chi_{55}&=&        \left(S_2 C_2 V_2 O_6+C_2 S_2
   O_2 V_6\right) \tau _{30}+\left(S_2 S_2 O_2
   O_6+C_2 C_2 V_2 V_6\right) \tau
   _{31}\   +\left(C_2 C_2 O_2 O_6+S_2 S_2
   V_2 V_6\right) \tau _{32}+\left(C_2 S_2 V_2
   O_6+S_2 C_2 O_2 V_6\right) \tau
   _{33}\nn\\  \chi_{56}&=&        \left(S_2 S_2 O_2 O_6+C_2 C_2
   V_2 V_6\right) \tau _{30}+\left(S_2 C_2 V_2
   O_6+C_2 S_2 O_2 V_6\right) \tau
   _{31}\   +\left(C_2 S_2 V_2 O_6+S_2 C_2
   O_2 V_6\right) \tau _{32}+\left(C_2 C_2 O_2
   O_6+S_2 S_2 V_2 V_6\right) \tau
   _{33}\nn\\  \chi_{57}&=&               \left(O_2 O_2 C_2 C_6+V_2
   V_2 S_2 S_6\right) \tau _{30}+\left(O_2 V_2
   S_2 C_6+V_2 O_2 C_2 S_6\right) \tau
   _{31}\   +\left(V_2 O_2 S_2 C_6+O_2 V_2
   C_2 S_6\right) \tau _{32}+\left(V_2 V_2 C_2
   C_6+O_2 O_2 S_2 S_6\right) \tau
   _{33}\nn\\  \chi_{58}&=&        \left(O_2 V_2 S_2 C_6+V_2 O_2
   C_2 S_6\right) \tau _{30}+\left(O_2 O_2 C_2
   C_6+V_2 V_2 S_2 S_6\right) \tau
   _{31}\   +\left(V_2 V_2 C_2 C_6+O_2 O_2
   S_2 S_6\right) \tau _{32}+\left(V_2 O_2 S_2
   C_6+O_2 V_2 C_2 S_6\right) \tau
   _{33}\nn\\  \chi_{59}&=&        \left(V_2 O_2 S_2 C_6+O_2 V_2
   C_2 S_6\right) \tau _{30}+\left(V_2 V_2 C_2
   C_6+O_2 O_2 S_2 S_6\right) \tau
   _{31}\   +\left(O_2 O_2 C_2 C_6+V_2 V_2
   S_2 S_6\right) \tau _{32}+\left(O_2 V_2 S_2
   C_6+V_2 O_2 C_2 S_6\right) \tau
   _{33}\nn\\  \chi_{60}&=&        \left(V_2 V_2 C_2 C_6+O_2 O_2
   S_2 S_6\right) \tau _{30}+\left(V_2 O_2 S_2
   C_6+O_2 V_2 C_2 S_6\right) \tau
   _{31}\   +\left(O_2 V_2 S_2 C_6+V_2 O_2
   C_2 S_6\right) \tau _{32}+\left(O_2 O_2 C_2
   C_6+V_2 V_2 S_2 S_6\right) \tau
   _{33}\nn\\  \chi_{61}&=&               \left(V_2 V_2 S_2 C_6+O_2
   O_2 C_2 S_6\right) \tau _{30}+\left(V_2 O_2
   C_2 C_6+O_2 V_2 S_2 S_6\right) \tau
   _{31}\   +\left(O_2 V_2 C_2 C_6+V_2 O_2
   S_2 S_6\right) \tau _{32}+\left(O_2 O_2 S_2
   C_6+V_2 V_2 C_2 S_6\right) \tau
   _{33}\nn\\  \chi_{62}&=&        \left(V_2 O_2 C_2 C_6+O_2 V_2
   S_2 S_6\right) \tau _{30}+\left(V_2 V_2 S_2
   C_6+O_2 O_2 C_2 S_6\right) \tau
   _{31}\   +\left(O_2 O_2 S_2 C_6+V_2 V_2
   C_2 S_6\right) \tau _{32}+\left(O_2 V_2 C_2
   C_6+V_2 O_2 S_2 S_6\right) \tau
   _{33}\nn\\  \chi_{63}&=&        \left(O_2 V_2 C_2 C_6+V_2 O_2
   S_2 S_6\right) \tau _{30}+\left(O_2 O_2 S_2
   C_6+V_2 V_2 C_2 S_6\right) \tau
   _{31}\   +\left(V_2 V_2 S_2 C_6+O_2 O_2
   C_2 S_6\right) \tau _{32}+\left(V_2 O_2 C_2
   C_6+O_2 V_2 S_2 S_6\right) \tau
   _{33}\nn\\  \chi_{64}&=&        \left(O_2 O_2 S_2 C_6+V_2 V_2
   C_2 S_6\right) \tau _{30}+\left(O_2 V_2 C_2
   C_6+V_2 O_2 S_2 S_6\right) \tau
   _{31}\   +\left(V_2 O_2 C_2 C_6+O_2 V_2
   S_2 S_6\right) \tau _{32}+\left(V_2 V_2 S_2
   C_6+O_2 O_2 C_2 S_6\right) \tau
   _{33}\nn
  \eea}}
\be~\ee


\end{document}